\documentclass[pdflatex,aps,twocolumn,superscriptaddress,prd,fleqn,showpacs,nofootinbib,preprintnumbers]{revtex4-1}
\usepackage{amssymb,amsmath,epsfig,color}
\usepackage{times,txfonts}
\usepackage{float} 
\usepackage[caption=false]{subfig} 

 \usepackage{ifpdf}
 \ifpdf
 \usepackage[pdftex]{hyperref}
 \else
 \usepackage[hypertex]{hyperref}
 \fi

 \hypersetup{
   pdftitle={},%
   pdfauthor={},%
   pdfsubject={},%
   pdfkeywords={},%
   pdfstartview={},%
   bookmarksopen=true, breaklinks=true, debug=true, %
   colorlinks=true, linkcolor=red, citecolor=blue, urlcolor=blue
 }

\def\to{\rightarrow}

\def\beq{\begin{equation}}
\def\eeq{\end{equation}}
\def\beeq{\begin{eqnarray}}
\def\eeeq{\end{eqnarray}}
\def\beal{\begin{align}}
\def\eeal{\end{align}}

\def\nn{\nonumber}
\def\b0{b_0}

\def\ID{1 \kern -.45 em 1}

\def\slash#1{\setbox0=\hbox{$#1$}               % set a box for #1
        \dimen0=\wd0                            % and get its size
        \setbox1=\hbox{/} \dimen1=\wd1          % get size of /
        \ifdim\dimen0>\dimen1                   % #1 is bigger
        \rlap{\hbox to \dimen0{\hfil/\hfil}}    % so center / in box
        #1                                      % and print #1
        \else              
        \rlap{\hbox to \dimen1{\hfil$#1$\hfil}} % so center #1
        /                                       % and print /
        \fi}                                    %

\begin{document}

\title{Double-Longitudinal Spin Asymmetry in Single-Inclusive Lepton Scattering at NLO}

\author{Patriz Hinderer}
\email{patriz.hinderer@uni-tuebingen.de}
\affiliation{Institute for Theoretical Physics, Universit\"{a}t T\"{u}bingen, Auf der Morgenstelle 14, D-72076 T\"{u}bingen, Germany}

\author{Marc Schlegel}
\email{marc.schlegel@uni-tuebingen.de}
\affiliation{Institute for Theoretical Physics, Universit\"{a}t T\"{u}bingen, Auf der Morgenstelle 14, D-72076 T\"{u}bingen, Germany}

\author{Werner Vogelsang}
\email{werner.vogelsang@uni-tuebingen.de}
\affiliation{Institute for Theoretical Physics, Universit\"{a}t T\"{u}bingen, Auf der Morgenstelle 14, D-72076 T\"{u}bingen, Germany}

%%%%%%%%%%%%%%%%%%%%%%%%%%%%%%%%%%%%%%%%%%%%%%%%%%%%%%%%%%%%
%%%%%%%%%%%%%%%%%%%%%%%%%%%       ABSTRACT      %%%%%%%%%%%%%%%%%%%%%%%
%%%%%%%%%%%%%%%%%%%%%%%%%%%%%%%%%%%%%%%%%%%%%%%%%%%%%%%%%%%%

\begin{abstract}
We calculate the double-spin asymmetries $A_{\mathrm{LL}}$ for the processes $\ell N\to h X$ and $\ell N\to {\mathrm{jet}}\, X$ 
at next-to-leading order accuracy in perturbative QCD. We compare our theoretical results for $A_{\mathrm{LL}}$ to data 
from the SLAC E155 experiment, finding only partially satisfactory agreement. We conclude that measurements 
of $A_{\mathrm{LL}}$ and the relevant polarized and unpolarized cross sections should be performed at the
present-day fixed-target lepton scattering experiments, as well as at a future electron ion collider, in order to 
verify our understanding of this process. We present predictions of the longitudinal double-spin 
asymmetries for these experiments.
\end{abstract}

\date{\today}

\maketitle

%%%%%%%%%%%%%%%%%%%%%%%%%%%%%%%%%%%%%%%%%%%%%%%%%%%%%%%%%%%%%%
%%%%%%%%%%%%%%%%%%%%%%%%%%%    INTRODUCTION     %%%%%%%%%%%%%%%%%%%%%%%%
%%%%%%%%%%%%%%%%%%%%%%%%%%%%%%%%%%%%%%%%%%%%%%%%%%%%%%%%%%%%%%

\section{Introduction}

The single-inclusive production of hadrons (or jets) with large transverse momenta in lepton-nucleon collisions, 
$\ell N\to h X$, has attracted much interest in recent years from both the 
experimental~\cite{E155,HERMESLL,HERMES,Hulse:2015caa,JLab} and 
the theoretical sides~\cite{Koike:2002gm,Kang:2011jw,Gamberg:2014eia,DSA,Lambda,LIR,GPM1,GPM2,GPM3,ourpaper,ourproc,Kang:2013wca,Kang:2013lga,JetNNLO,DAlesio:2017nrd}. The main reason for this interest is that $\ell N\to h X$
may prove to be particularly useful for obtaining a better understanding of transverse (nucleon) spin effects. 
As is well known, measurements for the related purely hadronic process $pp\to h X$ have revealed large transverse 
single-spin asymmetries $A_{\mathrm{N}}$~\cite{Aidala:2012mv}, and the understanding of these large effects 
remains to pose a major challenge to theory. Since the process $\ell N\to h X$ is generally simpler to analyze 
theoretically, it is hoped that its transverse single-spin asymmetry will help us to identify the origins of the 
large effects observed in hadronic scattering. 

The basis for the theoretical description of single-inclusive processes is collinear factorization in perturbative QCD (pQCD). 
In many instances, next-to-leading order (NLO) corrections, or even corrections beyond NLO, are found to be
sizable for single-inclusive scattering~\cite{barbara,deflo,Uebler}. Therefore, in order to be able to reliably confront data and theory
for the single-transverse spin asymmetry for $\ell N\to h X$, it is crucial to have the NLO corrections for the cross
sections entering the asymmetry. The recent study~\cite{Gamberg:2014eia} of $A_{\mathrm{N}}$ in $\ell p \to hX$ 
at leading order (LO) suggests the presence of sizable NLO corrections: comparing to recent HERMES data~\cite{HERMES} 
it was found that the LO theoretical prediction lies significantly higher than the data. 

To compute the NLO corrections for the single-transverse spin asymmetry is a very complex task, however. The
asymmetry is power-suppressed in QCD and involves twist-3 three-parton correlations. Higher-order
calculations for higher-twist single-spin observables are notoriously difficult in pQCD and still relatively 
scarce~\cite{DYNLO}. As a first step towards an NLO calculation for the spin asymmetries in $\ell p \to hX$
(or $\ell p \to {\mathrm{jet}}\,X$), we have recently computed the NLO results for the respective spin-averaged
cross sections~\cite{ourpaper,ourproc} that constitute the denominators of the asymmetries. 
Indeed, large NLO corrections were found for the various kinematic
regimes of interest. We note that very recently even the next-to-next-to leading order calculation for the 
spin-averaged cross section for $\ell p \to {\mathrm{jet}}\,X$ was presented~\cite{JetNNLO}. 

Unfortunately, so far no experimental data on the unpolarized cross sections exist that would allow for a 
comparison between theory and data. On the other hand, some data from SLAC~\cite{E155} and 
HERMES~\cite{HERMESLL} are available on the {\it double-longitudinal} spin asymmetry $A_{\mathrm{LL}}$, 
measured by scattering a longitudinally polarized lepton off a longitudinally polarized target. This asymmetry 
is of leading power in QCD and can hence be analyzed with standard techniques in pQCD. 
Therefore, prior to studying the more complex case of $A_{\mathrm{N}}$, we investigate
$A_{\mathrm{LL}}$ at NLO in this paper. 

For the single-inclusive processes $\ell p \to hX$ and $\ell p \to {\mathrm{jet}}\,X$ the scattered
lepton is not explicitly detected. As a result, there are contributions to the cross sections for
which the incident lepton emits an almost real photon, followed by a hard photoproduction scattering
process. Such contributions are formally NLO, but they are enhanced by the photon propagator. 
In Refs.~\cite{Afanasev:1996mj} and~\cite{ep7} (at NLO, including resolved-photon contributions) 
the asymmetry $A_{\mathrm{LL}}$ for $\ell p \to hX$ was investigated under the approximation that 
the scattering is entirely dominated by the exchange of such quasi-real photons. However, as we 
found in~\cite{ourpaper}, for the spin-averaged cross section this assumption is valid only in very 
limited kinematical regions. In general, a full NLO calculation is needed, for which the quasi-real 
photon contribution is just one among several. In this paper we investigate in how far the assumption 
of dominant exchange of quasi-real photons is justified for the double-longitudinally polarized cross 
section and for $A_{\mathrm{LL}}$. In this context we also present new comparisons to the E155 data \cite{E155},
on the basis of a full NLO calculation. 

Our paper is structured as follows. In Sec.~\ref{nlocalc} we present our NLO calculations for the longitudinally 
polarized 
cross sections for $\ell N\to h X$ and $\ell N\to \mathrm{jet}\, X$. Section~\ref{E155} presents 
a comparison to the E155 data. Section~\ref{pheno} presents numerical predictions for the NLO double-spin asymmetry 
$A_{\mathrm{LL}}$ to be expected at various other fixed-target experiments and at a future Electron Ion Collider (EIC). 
Finally, we summarize our results in Sec.~\ref{concl}.
 
%%%%%%%%%%%%%%%%%%%%%%%%%%%%%%%%%%%%%%%%%%%%%%%%%%%%%%%%%%%%
%%%%%%%%%%%%%%%%%%%%%%%%%%%%%%%%% Main Body  %%%%% %%%%%%%%%%%%%%%%
%%%%%%%%%%%%%%%%%%%%%%%%%%%%%%%%%%%%%%%%%%%%%%%%%%%%%%%%%%%%

\section{NLO calculation \label{nlocalc}}

\subsection{Single-Inclusive Hadron Production}

In this section we briefly present our derivation of the analytical NLO results for the
processes $\ell N\to h X$ and $\ell N\to {\mathrm{jet}}\, X$ with longitudinally polarized
initial particles. We will closely follow our previous paper \cite{ourpaper} in which we
computed the corresponding unpolarized NLO cross sections. We will be brief and highlight
only the differences arising for longitudinal polarization. We refer 
the reader to Ref.~\cite{ourpaper} for details concerning the calculation.

The transverse momentum of the produced hadron sets a hard scale, so that 
perturbative methods may be used for treating the cross sections. We first consider $\ell(l) + N(P)\rightarrow h (P_h)+X$, 
where we have introduced our notation for the four-momenta. We define the Mandelstam variables 
as $S=(P+l)^2$, $T=(P-P_h)^2$ and $U=(l-P_h)^2$. Furthermore, we denote the energy of the detected hadron by
$E_h$ and its three-momentum by $\vec{P}_h$. The momenta of the incoming parton, $k^\mu$, and 
of the fragmenting parton, $p^\mu$, which appear in the calculation of the partonic cross sections, are approximated 
as $k^\mu\simeq x P^\mu$ and $p^\mu \simeq P_h^\mu /z$, respectively. It is then convenient to work with the partonic 
Mandelstam variables 
\beeq\label{stu}
s=(k+l)^2\approx xS,\;\, t=(k-p)^2\approx \frac{x}{z}T,\;\,u=(l-p)^2\approx\frac{U}{z}\,.
\eeeq
We will consider the following difference of cross sections:
\beeq
\Delta\sigma & \equiv& \frac{1}{2}\left[E_h \frac{d^3\sigma^{\ell N\to h X}(S_L=+1,\lambda_e=+1)}{d^3P_h}\right.\nonumber\\
&& - \left.E_h \frac{d^3\sigma^{\ell N\to h X}(S_L=+1,\lambda_e=-1)}{d^3P_h}\right]\,.\label{DefAsym}
\eeeq
In this expression $S_L$ and $\lambda_{\ell}$ denote the helicities of the nucleon and the lepton, respectively. This choice of difference between polarized cross sections corresponds to the numerator of the longitudinal double-spin asymmetry $A_{\parallel}$ that was measured by the E155 experiment~\cite{E155}. 

The general form of the factorized polarized cross section for inclusive hadron production process is then 
\beeq 
\Delta \sigma&=&\frac{1}{S}\sum_{i, f}
\int_{0}^1 \frac{dx}{x}\int_{0}^1 \frac{dz}{z^2} \, \Delta f^{i/N}(x,\mu)\nonumber\\[2mm]
&\times & D^{h/f}(z,\mu)\;\Delta\hat{\sigma}^{i\to f}(s,t,u,\mu)\;,
\label{invariantcs1}
\eeeq
where $\Delta f^{i/N}(x,\mu)$ is the helicity parton distribution function for the incoming parton $i$ in the nucleon $N$ 
and $D^{h/f}(z,\mu)$ the fragmentation function for parton $f$ fragmenting into hadron $h$, 
both evaluated at a factorization scale $\mu$. As in Ref.~\cite{ourpaper} we choose the factorization scales to be the same for the initial
and the final state, and also equal to the renormalization scale. In Eq.~(\ref{invariantcs1}), $\Delta \hat{\sigma}^{i\to f}$ 
is the difference of longitudinally polarized cross sections for the lepton-parton scattering process $\ell+ i\to f+x$, with $x$ an unobserved final state. This difference is defined in analogy with that in Eq.~(\ref{DefAsym}) 
The sum in Eq.~(\ref{invariantcs1}) runs over the different species of partons, quarks, gluons and
antiquarks.  We note that the expression in Eq.~(\ref{invariantcs1}) holds up to corrections that are suppressed
by inverse powers of the produced hadron's transverse momentum $P_{h\perp}$. 

It is convenient to rewrite the $x$- and $z$-integrals in Eq.~(\ref{invariantcs1}) in terms of new 
variables $v\equiv 1+t/s$ and $w\equiv-u/(s+t)$. Using~(\ref{stu}), we have
\beq
x=\frac{1-v}{vw}\frac{U}{T}\,,\;\;z=\frac{-T}{(1-v)S}\,,
\eeq
and Eq.~(\ref{invariantcs1}) becomes
\beeq 
\Delta\sigma&=&\left(\frac{-U}{S^2}\right)\sum_{i, f}
\int_{\frac{U}{T+U}}^{1+\frac{T}{S}} \frac{dv}{v(1-v)}\int_{\tfrac{1-v}{v}\tfrac{U}{T}}^1 \frac{dw}{w^2} 
\nonumber\\[2mm]
& \times & \frac{\Delta f^{i/N}(x,\mu)}{x}\frac{D^{h/f}(z,\mu)}{z^2} \;\Delta\hat{\sigma}^{i\to f}(v,w,\mu)\;,
\label{Trafox}
\eeeq
where $x=\tfrac{1-v}{vw}\tfrac{U}{T},\,z=\tfrac{-T}{(1-v)S}$.
For ease of notation, we have kept the symbol $\Delta\hat{\sigma}^{i\to f}$ also for the polarized cross section when
expressed in terms of the new variables. We note that the invariant mass of the unobserved recoiling 
final state is given by $s+t+u=s v (1-w)$. 

The partonic polarized cross sections $\Delta\hat{\sigma}^{i\to f}$ in Eq.~(\ref{Trafox}) can be calculated in QCD perturbation
theory. One may write their expansions in the strong coupling $\alpha_s$ as
\beq
\Delta\hat{\sigma}^{i\to f}\,=\,\Delta\hat{\sigma}^{i\to f}_{\mathrm{LO}} + \frac{\alpha_s}{\pi}\,\Delta\hat{\sigma}^{i\to f}_{\mathrm{NLO}}+
{\cal O}(\alpha_s^2)\,.
\eeq
As explained in detail in Ref.~\cite{ourpaper}, there are contributions to the NLO cross section for which the
photon exchanged between lepton and quark is almost real. These contributions are typically sizable. In fact 
they diverge when the mass of the lepton tends to zero. Expanding the partonic cross section in the (small) 
lepton mass $m_\ell$, one finds the structure
\begin{eqnarray}
\Delta \hat{\sigma}^{i\to f}_{\mathrm{NLO}}(v,w,m_\ell,\mu) &=& \Delta \hat{\sigma}^{i\to f}_{\mathrm{log}}(v,w)\; 
\log(m_\ell/\mu) +\nonumber\\[2mm]
&& \Delta \hat{\sigma}^{i\to f}_{0}(v,w,\mu/s)+\mathcal{O}(m_\ell^2\log(m_\ell)) ,\label{massive}
\end{eqnarray}
which exhibits the ``mass singularity'' as $m_\ell\to 0$. The scale $\mu$ is arbitrary and cancels between 
the first two pieces. The logarithmic part in~(\ref{massive}) in a sense opens up new partonic channels at NLO,
since it arises from configurations where the lepton radiates the photon collinearly, which subsequently 
participates as an initial particle in a photoproduction scattering process. Dropping all terms that vanish for $m_\ell=0$ 
we arrive at the following form of the NLO cross section:
\beeq 
\Delta \sigma&=&\left(\frac{-U}{S^2}\right)\sum_{i, f}
\int_{\frac{U}{T+U}}^{1+\frac{T}{S}} \frac{dv}{v(1-v)}\int_{\tfrac{1-v}{v}\tfrac{U}{T}}^1 \frac{dw}{w^2} 
\nonumber\\[2mm]
& \times & \frac{\Delta f^{i/N}(x,\mu)}{x}\frac{D^{h/f}(z,\mu)}{z^2} \;\left[\Delta \hat{\sigma}^{i\to f}_{\mathrm{LO}} (v)+ 
\frac{\alpha_s(\mu)}{\pi}\,\Delta \hat{\sigma}^{i\to f}_{\mathrm{NLO}}(v,w,\mu)\right.\nn\\[2mm]
&+&\Delta f^{\gamma/\ell}\left(\tfrac{1-v}{1-vw},\mu\right)\,
\frac{\alpha_s(\mu)}{\pi}\,\Delta \hat{\sigma}_{\mathrm{LO}}^{\gamma i\to f}(v,w)
\Bigg]\;,
\label{Trafox1}
\eeeq
where 
\begin{eqnarray}
\Delta f^{\gamma/\ell}(y,\mu)& \equiv & \frac{\alpha_{\mathrm{em}}}{2\pi}\;\Delta P_{\gamma\ell}(y)\; \log \left(\frac{\mu^2}{y^2 m_\ell^2}\right) +
\mathcal{O}(\alpha_{\mathrm{em}}^2)\label{fWWren}
\end{eqnarray}
is the polarized ``photon-in-lepton'' distribution, with $\alpha_{\mathrm{em}}$ the fine structure constant. 
It can be calculated perturbatively as discussed in Ref.~\cite{ourpaper}
and involves the polarized lepton-photon splitting function $\Delta P_{\gamma/\ell}(y)=2-y$. The $\Delta \hat{\sigma}_{\mathrm{LO}}^
{\gamma i\to f}$ are the spin-dependent lowest-order scattering cross sections for $\gamma+i\to f+x$, computed
with real incoming photons. They will be given below. We stress again that Eq.~(\ref{Trafox1}) is exact up to 
terms that vanish as $m_\ell\to 0$. The same result may be obtained in a calculation that treats the lepton as massless 
from the beginning. The ensuing collinear divergence may then be absorbed into a ``bare'' photon-in-lepton 
distribution and is canceled in this way~\cite{ourpaper}. 

For the LO partonic cross section in~(\ref{Trafox1}), present only for the channel $q\to q$ with an incoming quark 
that also fragments, one finds
\begin{equation}
\Delta\hat{\sigma}^{q\to q}_{\mathrm{LO}} = 
2 \alpha_{\mathrm{em}}^2 e_q^2\frac{1}{sv}\; \frac{1-v^2}{(1-v)^2}\; \delta(1-w)\; ,
\label{CSLOvw1eps}
\end{equation}
where $e_q$ is the quark's fractional charge.
 
The NLO terms may be computed using the techniques discussed in~\cite{ourpaper}. 
The only new technical aspect concerns the use of the Dirac matrix $\gamma_5$ and the 
Levi-Civita tensor $\epsilon^{\mu\nu\rho\sigma}$ in dimensional regularization, 
which appear in the projections onto helicity states for the incoming particles. We use 
the 't Hooft-Veltman-Breitenlohner-Maison (HVBM) scheme \cite{tHooftVeltman,BreitenlohnerMaison} 
throughout our calculation. For details of the application of these scheme in NLO calculations of
single-inclusive cross sections we refer the reader to \cite{ep7a}. As is well-known, the HVBM scheme
produces spurious terms that violate helicity conservation at the quark-gluon vertex~\cite{wv2loop}. 
This feature manifests itself in a spin-dependent splitting function $\Delta P_{qq}(y)$ in $d= 4-2\epsilon$
dimensions that differs from the spin-averaged one, $\Delta P_{qq}(y)=P_{qq}(y)+\varepsilon \,4 C_F(1-y)$.
This may be corrected by a finite subtraction in the process of factorization of collinear singularities.
This is the standard choice made in the literature and is also in accordance with all modern
sets of NLO helicity parton distribution functions. 

For the NLO term in the $q\to q$ channel we find
\begin{eqnarray}
\Delta \hat{\sigma}_{\mathrm{NLO}}^{q\to q}(v,w,\mu)&=&\frac{\alpha_{\mathrm{em}}^2 e_q^2C_F}{svw}\Bigg[
\Delta A_0^{q\to q} \,\delta(1-w)\nonumber\\
&+& \Delta A_1^{q\to q} \left(\frac{\log(1-w)}{1-w}\right)_+\nn\\[2mm]
&+& \frac{1}{(1-w)_+}\Bigg\{\Delta B_{1}^{q\to q}\log\left(\frac{1-v}{v(1-v(1-w))}\right)\nonumber\\[2mm]
&+&\Delta B_{2}^{q\to q}\log(1-v(1-w))+\Delta B_{3}^{q\to q}\log\left(\frac{sv^2}{\mu^2}\right)\Bigg\}\nn\\[2mm]
&+&\Delta C_1^{q\to q}\log(v(1-w))+\Delta C_2^{q\to q}\log\left(\frac{(1-v)w}{1-vw}\right)\nonumber\\[2mm]
&+& \Delta C_3^{q\to q}\log\left(\frac{1-v}{(1-vw)(1-v(1-w))}\right)\nonumber\\[2mm]
&+& \Delta C_4^{q\to q}\log\left(1-v(1-w)\right)\nonumber\\[2mm]
&+&\Delta C_5^{q\to q}\log\left(\frac{s}{\mu^2}\right)+\Delta C_6^{q\to q}\Bigg]\,,\label{Resq2qNLOreal1}
\end{eqnarray}
where $C_F=4/3$.The coefficients $\Delta A_i^{q\to q}$, $\Delta B_i^{q\to q}$, $\Delta C^{q\to q}_{i}$ are functions of 
$v$ and $w$ and may be found in the Appendix. Equation (\ref{Resq2qNLOreal1}) contains the usual ``plus distributions''
defined as
\beq
\int_0^1dw\, f(w)\left[g(w)\right]_+  = \int_0^1dw\, \left[f(w)-f(1)\right]g(w) \;.\label{Plus}
\eeq
For the channels $q\to g$ and $g\to q$ we find the simpler expressions
\begin{eqnarray}
\Delta \hat{\sigma}_{\mathrm{NLO}}^{q\to g}(v,w,\mu) & = & \frac{\alpha_{\mathrm{em}}^2 e_q^2C_F}{svw}
\Bigg[\Delta C_1^{q\to g}\log(1-v(1-w))\nonumber\\[2mm]
&+&\Delta  C_2^{q\to g}\log\left(\frac{1-v}{(1-vw)(1-v(1-w))}\right)\nonumber\\[2mm]
&+&\Delta C_3^{q\to g}\log\left(\frac{v(1-w)s}{\mu^2}\right)+\Delta C_4^{q\to g}\Bigg]\,,
\label{Resq2gNLOreal1}
\end{eqnarray}
and
\begin{eqnarray}
\Delta\hat{\sigma}_{\mathrm{NLO}}^{g\to q}(v,w,\mu) & = & 
\frac{\alpha_{\mathrm{em}}^2 e_q^2T_R}{svw}
\Bigg[\Delta C_1^{g\to q}\log\left(\frac{(1-v)w}{1-vw}\right)\nonumber\\[2mm]
&+&\Delta C_2^{g\to q}\log\left(\frac{v(1-w)s}{\mu^2}\right)+\Delta C_3^{g\to q}\Bigg]\,,
\label{Resg2qNLOreal1}
\end{eqnarray}
where $T_R=1/2$. 
The coefficients $\Delta C^{q\to g}_{i}$ and $\Delta C^{g\to q}_{i}$ are again given in the Appendix.
We finally list the spin-dependent partonic cross sections for the photon-initiated channels:
\begin{eqnarray}\label{WWsigs}
\Delta \hat{\sigma}_{\mathrm{LO}}^{\gamma q\to q}(v,w)&=&\frac{2\pi\,C_F \alpha_{\mathrm{em}}e_q^2}{s(1-v)}
\frac{1-v^2 w^2}{vw}\,,\nn\\[2mm]
\Delta \hat{\sigma}_{\mathrm{LO}}^{\gamma q\to g}(v,w)&=&\frac{2\pi\,C_F \alpha_{\mathrm{em}}e_q^2}{s(1-v)}
\frac{vw(2-vw)}{1-vw}\,,\nn\\[2mm]
\Delta \hat{\sigma}_{\mathrm{LO}}^{\gamma g\to q}(v,w)&=&-
\frac{2\pi\,T_R\alpha_{\mathrm{em}}e_q^2}{s(1-v)}\frac{v^2 w^2 + (1-vw)^2}{vw(1-vw)}\,.
\end{eqnarray}

\subsection{Single-inclusive Jet Production \label{jetpro}}

Having computed the NLO polarized cross section for inclusive hadron production the extension to single inclusive jet 
production is relatively straightforward, using the techniques of Refs.~\cite{Jager:2004jh,Mukherjee:2012uz,Kaufmann:2014nda}.
The spin-dependent cross section for $\ell N\to {\mathrm{jet}}\,X$ may be written as
\begin{eqnarray}
\Delta \sigma^{\ell N\to \mathrm{jet}X}& = &
\frac{1}{S}\sum_i
\int_{\frac{-U}{S+T}}^1\frac{dw}{w}\,\Delta f^{i/N}\left(x=\tfrac{-U}{w(S+T)},\mu\right)\nn\\[2mm]
&\times&\left[\Delta \hat{\sigma}^{i\to \mathrm{jet}}_{\mathrm{incl.\; parton}}\left(v=1+\frac{T}{S},w,\mu\right)\right.\nonumber\\
&+&\left.\Delta \hat{\sigma}^{i\to \mathrm{jet}}_R\left(v=1+\frac{T}{S},w,\mu;R\right)\right] \label{jet}\,.
\end{eqnarray}
As indicated, the partonic cross section is the sum of two contributions. The first contains
inclusive-parton cross sections $\Delta \hat{\sigma}^{i\to \mathrm{jet}}_{\mathrm{incl.\; parton}}$.
This part of the cross section is obtained from~(\ref{Trafox1}) by setting the fragmentation functions to $\delta(1-z)$. 
Explicitly, we have 
\begin{eqnarray}
\Delta \hat{\sigma}^{i\to \mathrm{jet}}_{\mathrm{incl.\; parton}}\left(v,w,\mu\right) &=&
\sum_f \left[ \Delta \hat{\sigma}^{i\to f}_{\mathrm{LO}} (v)+
\frac{\alpha_s(\mu)}{\pi}\,\Delta \hat{\sigma}^{i\to f}_{\mathrm{NLO}}(v,w,\mu)\right. \nn\\[2mm]
&&\hspace*{-1cm}\left.+\,\Delta f^{\gamma/\ell}\left(\tfrac{1-v}{1-vw},\mu\right)\,
\frac{\alpha_s(\mu)}{\pi}\,\Delta \hat{\sigma}_{\mathrm{LO}}^{\gamma i\to f}(v,w)\right],
\end{eqnarray}
where $v=1+T/S$. The inclusive-parton cross section has been integrated over the full phase space of the 
unobserved final-state particles, keeping the momentum of the observed particle fixed. This is not appropriate 
for an NLO jet cross section for which two final-state particles may jointly form the jet. As shown in 
Refs.~\cite{Jager:2004jh,Mukherjee:2012uz,Kaufmann:2014nda}, one can correct for this by
introducing a subtraction piece. This piece is represented by the term involving $\Delta \hat{\sigma}^{i\to \mathrm{jet}}_R$ 
in Eq.~(\ref{jet}). As discussed further in~\cite{Jager:2004jh,Mukherjee:2012uz,Kaufmann:2014nda} one may 
determine the $\Delta \hat{\sigma}^{i\to \mathrm{jet}}_R$ analytically if one assumes
that the jet size parameter $R$ is relatively small. Without going into further detail we just quote the final results
relevant for the spin-dependent cross section for $\ell N\to {\mathrm{jet}}\,X$:
\begin{eqnarray}
\Delta \hat{\sigma}^{q\to \mathrm{jet}}_R\left(v,w,\mu;R\right)&=& -\frac{\alpha_s(\mu)}{\pi}\frac{C_F \alpha_{\mathrm{em}}^2 e_q^2}{svw}\,
\Delta {\cal H}^{q\to q}(v,w)\nonumber\\
&&\hspace{-2.5cm}\times \Bigg[A_0^{\mathrm{jet}}(v;R)\,\delta(1-w)+A_1^{\mathrm{jet}}(v,w)\,\left(\frac{\log(1-w)}{1-w}\right)_++\nonumber\\
&&\hspace{-2.16cm}B_1^{\mathrm{jet}}(v,w;R)\,\frac{1}{(1-w)_+}+C_1^{\mathrm{jet}}(v,w;R)\Bigg]+\mathcal{O}(\alpha_s^2),\nonumber\\
\Delta \hat{\sigma}^{g\to \mathrm{jet}}_R\left(v,w,\mu;R\right)&=& \mathcal{O}(\alpha_s^2).\label{jetPartCS}
\end{eqnarray}
The coefficients $A_0^{\mathrm{jet}}$, $A_1^{\mathrm{jet}}$, $B_1^{\mathrm{jet}}$, $C_1^{\mathrm{jet}}$ are given in the Appendix. 
They show that the NLO jet cross section has the form ${\cal A}\log(R)+{\cal B}+{\cal O}(R^2)$. The coefficient $A_0^{\mathrm{jet}}$ 
depends on the jet algorithm used to define the jet parameter $R$. The result given in the Appendix refers to the 
anti-$k_T$ algorithm~\cite{Cacciari:2008gp}. The hard-scattering function in~(\ref{jetPartCS}) is related to the Born cross
section in Eq.~(\ref{CSLOvw1eps}):
\begin{equation}
\Delta {\cal H}^{q\to q}(v,w)=\frac{1-v^{\prime 2}}{(1-v^{\prime})^2}\Bigg|_{v^\prime=vw/(1-v(1-w))}.
\end{equation}
We note that for the unpolarized cross section the same coefficients $A_0^{\mathrm{jet}}$, $A_1^{\mathrm{jet}}$, $B_1^{\mathrm{jet}}$, $C_1^{\mathrm{jet}}$ appear, with however the hard part
\begin{equation}
\Delta {\cal H}^{q\to q}(v,w) \to {\cal H}^{q\to q}=\frac{1+v^{\prime 2}}{(1-v^{\prime})^2}\Bigg|_{v^\prime=vw/(1-v(1-w))}.\label{jetUUR}
\end{equation}

\section{Phenomenological Results \label{E155}}

%%%%%%%%%%%%%%%%%%%%%%%%%%%%%%%%%%%%%%%%%%%%%%%%%%%%%%%%%%%%%
\begin{figure*}[htb]
\centering
\subfloat[]{\includegraphics[width=0.5\textwidth,angle=0]{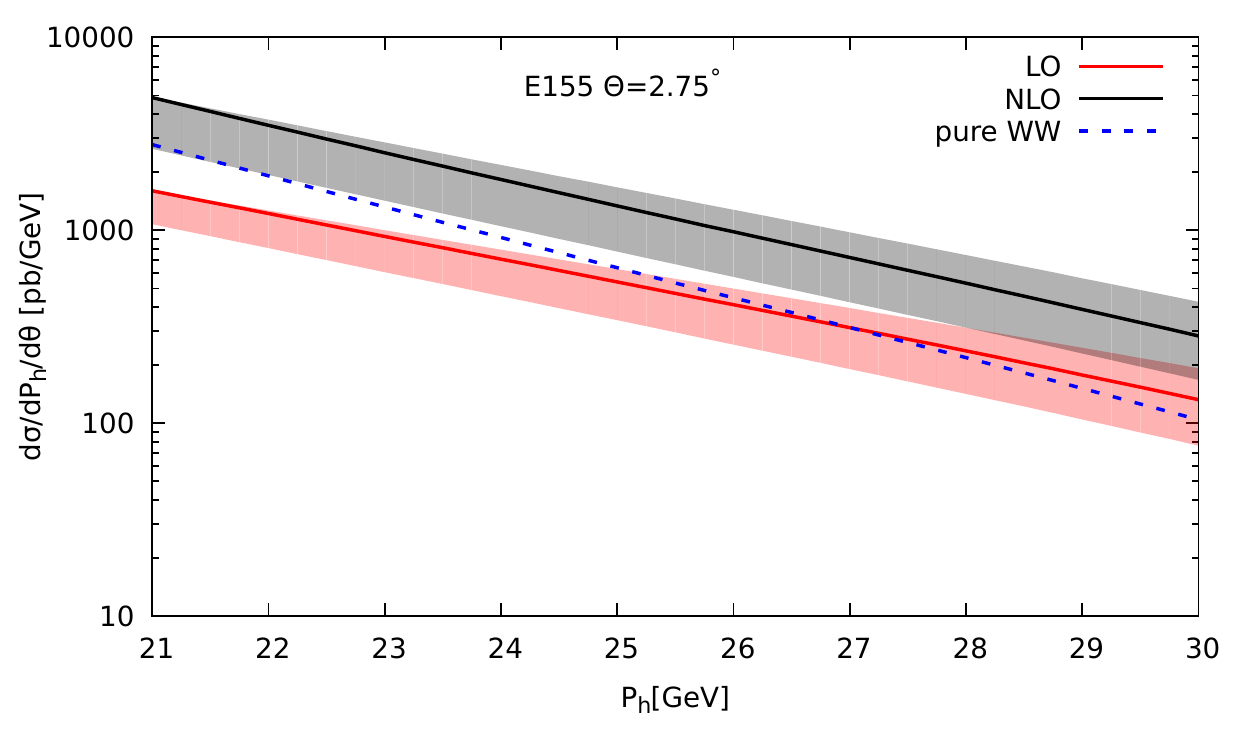}\label{fig:UUp2pip275}}
\subfloat[]{\includegraphics[width=0.5\textwidth,angle=0]{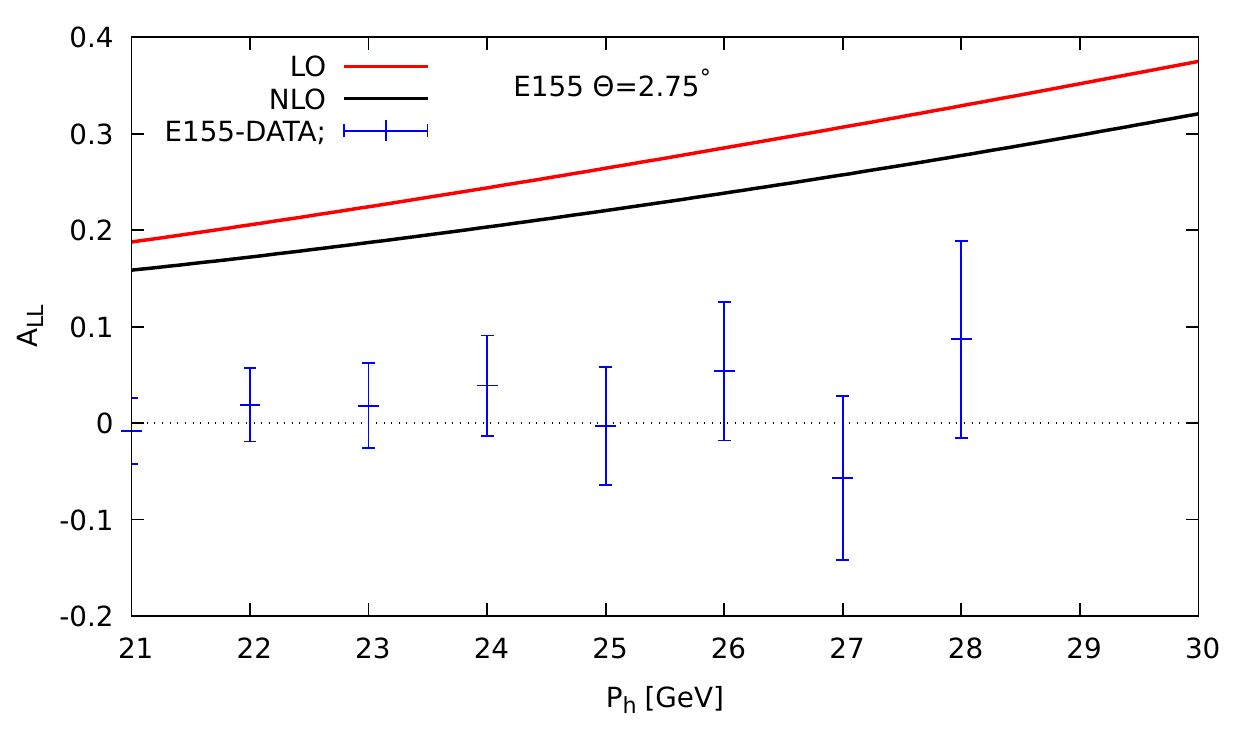}\label{fig:ALLp2pip275}}
\\
\subfloat[]{\includegraphics[width=0.5\textwidth,angle=0]{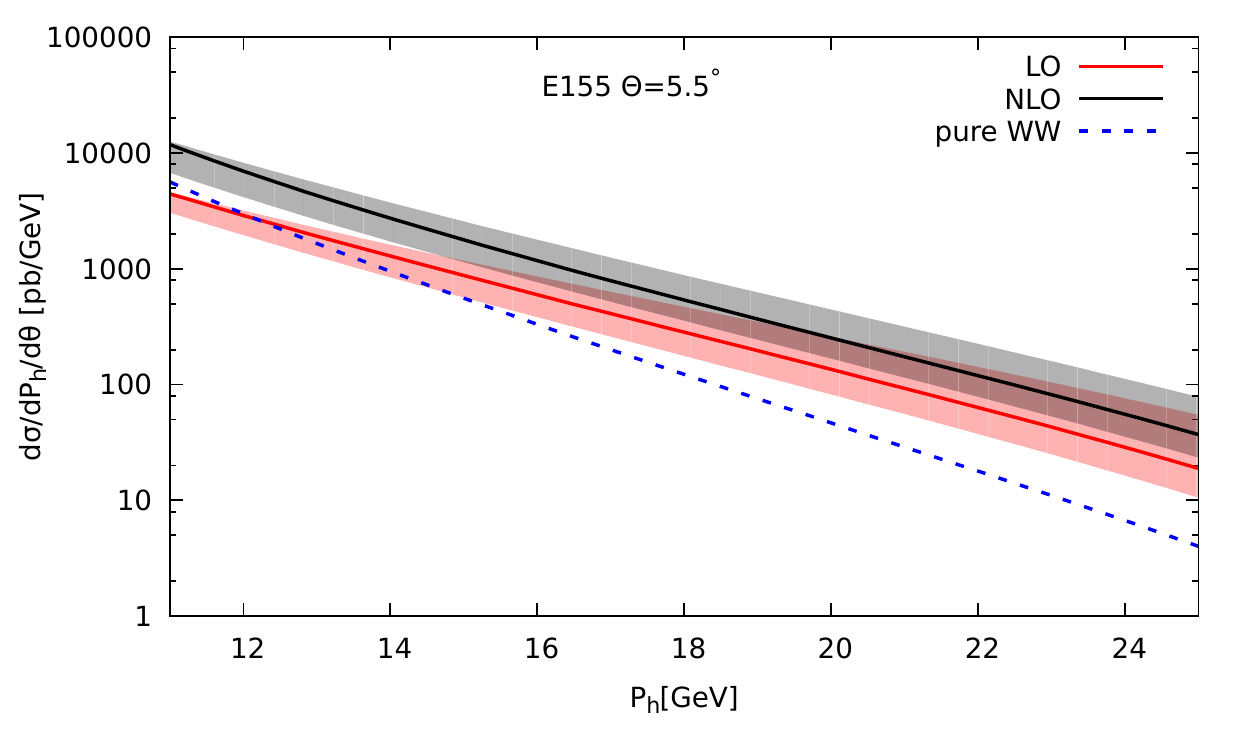}\label{fig:UUp2pip55}}
\subfloat[]{\includegraphics[width=0.5\textwidth,angle=0]{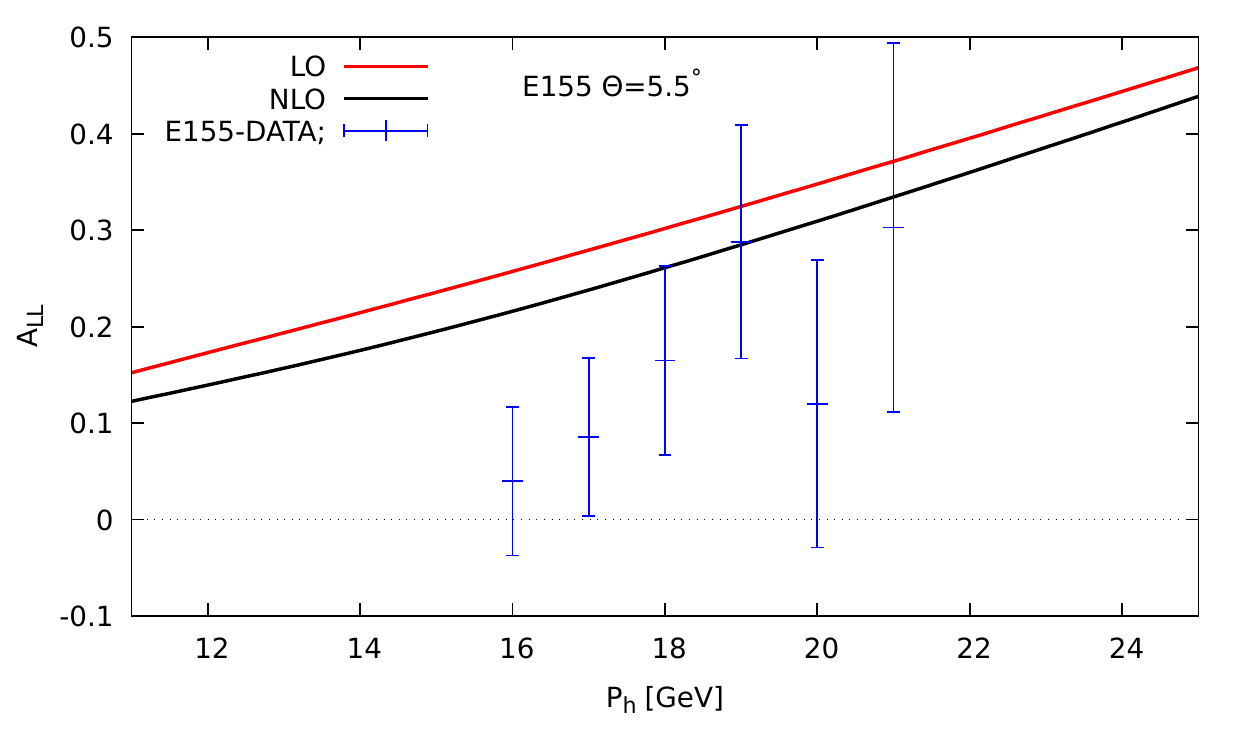}\label{fig:ALLp2pip55}}
\caption{Unpolarized cross sections ((a) and (c)) and longitudinal double-spin asymmetries 
$A_{\mathrm{LL}}$ for $ep\to \pi^+ X$, at scattering angle $\theta=2.75^{\circ}$ (upper panel) and $\theta=5.5^{\circ}$ (lower panel), respectively.
We show LO and NLO results. The data are from E155~\ref{E155}. The dashed line shows the 
pure Weizs\"{a}cker-Williams contributions by quasi-real photons. The bands in (a) and (c) represent the scale
variation $1\,\mathrm{GeV}<\mu < 2 P_{h\perp}$.}
\end{figure*}
%%%%%%%%%%%%%%%%%%%%%%%%%%%%%%%%%%%%%%%%%%%%%%%%%%%%%%%%%%%%%

In this section we present numerical estimates for the longitudinal double-spin asymmetry $A_{\mathrm{LL}}$ for 
the kinematical setup of the SLAC E155 experiment \cite{E155}. Although HERMES also reports a measurement of $A_{\mathrm{LL}}$ \cite{HERMESLL} in $ep\to hX$, we cannot compare to their data. The reason is that the HERMES data are
not fully single-inclusive but were taken with the requirement that the scattered electron not be seen within
the detector acceptance. This is different from a fully single-inclusive measurement for which the electron 
may be in any region of phase space. Also, hadron transverse momenta are typically very low for the HERMES
data. 

E155 used an electron beam with energy $E=48.35\,\mathrm{GeV}$ scattering off proton or deuteron targets. 
The experiment measured the double-longitudinal spin asymmetry 
\begin{equation}
A^{\ell N\to hX}_{{\mathrm{LL}}}(S,T,U)\equiv \frac{\Delta \sigma^{\ell N\to hX}_\mathrm{NLO}(S,T,U)}{\sigma^{\ell N\to hX}_{\mathrm{NLO}}(S,T,U)}.\label{ALL}
\end{equation}
at two scattering angles, $\theta=2.75^\circ$ and $\theta=5.5^\circ$, defined in the laboratory (target rest) frame, relative to the direction of the incident lepton beam. 
Hadrons with momenta $10\,\mathrm{GeV} \leq |\vec{P}_h| \leq 29\,\mathrm{GeV}$ were accepted, 
again defined in the laboratory frame. The asymmetry $A^{\ell N\to hX}_{{\mathrm{LL}}}$ was measured
as a function of $|\vec{P}_h|$. Data were presented for identified pions and also for unidentified charged
hadrons.

In order to compute $A^{\ell N\to hX}_{{\mathrm{LL}}}$ at NLO we use Eq.~(\ref{Trafox1}) for the spin-dependent cross 
section, accompanied by the results in Eq.~(25) of Ref.~\cite{ourpaper} for the spin-averaged one. In the 
target rest frame, neglecting the mass of the produced hadron, we have
\begin{eqnarray}
S & = & (P+l)^2 \,=\, 2ME+M^2\,,\nonumber\\
T & = & (P-P_h)^2 \,=\,M^2-2M|\vec{P}_h|\,,\nonumber\\
U & = & (l-P_h)^2 \,=\,-2E|\vec{P}_h|(1-\cos\theta)\,,\label{MandelstamE155}
\end{eqnarray}
where $M$ is the proton mass. We note that we find a rather strong decrease of our results 
(at the level of about 10\%) if we drop the $M^2$ terms in (\ref{MandelstamE155}). This 
is to be understood from the relatively modest beam energy and the forward kinematics.
In principle we should include the full set of target mass corrections which, however, 
is beyond the scope of this article. 

We note that the transverse hadron momentum is given in the rest-frame variables by 
$|\vec{P}_{h\perp}|=|\vec{P}_h|\sin(\theta)$. Since the transverse momentum sets the hard scale for the process
we demand for our calculations that $|\vec{P}_{h\perp}|\geq 1$~GeV. For the scattering angle $\theta=2.75^\circ$ 
this corresponds to a lower bound $|\vec{P}_h|\geq 20.9\,\mathrm{GeV}$ whereas for $\theta=5.5^\circ$ we 
have $|\vec{P}_h|\geq 10.5\,\mathrm{GeV}$. Conversely, for $\theta=2.75^\circ$ the data extend to 
$|\vec{P}_h|=29\,\mathrm{GeV}$, corresponding to transverse momenta 
$|\vec{P}_{h\perp}|\leq 1.4\,\mathrm{GeV}$. For $\theta=5.5^\circ$ the maximal hadron momentum in 
E155 is about $|\vec{P}_h|=24\,\mathrm{GeV}$, yielding $|\vec{P}_{h\perp}|\leq 2.3\,\mathrm{GeV}$.
%%%%%%%%%%%%%%%%%%%%%%%%%%%%%%%%%%%%%%%%%%%%%%%%%%%%%%%%%%%%%
\begin{figure*}[htb]
\centering
\subfloat[]{\includegraphics[width=0.5\textwidth,angle=0]{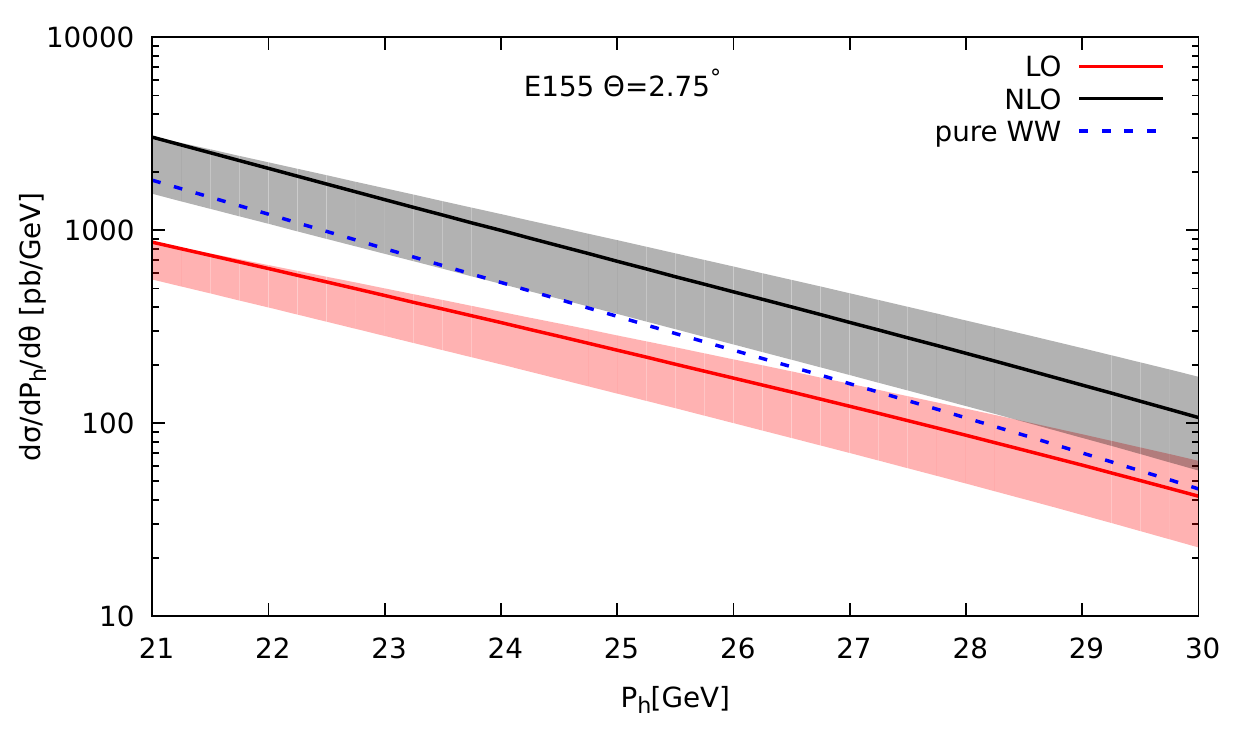}\label{fig:UUp2pim275}}
\subfloat[]{\includegraphics[width=0.5\textwidth,angle=0]{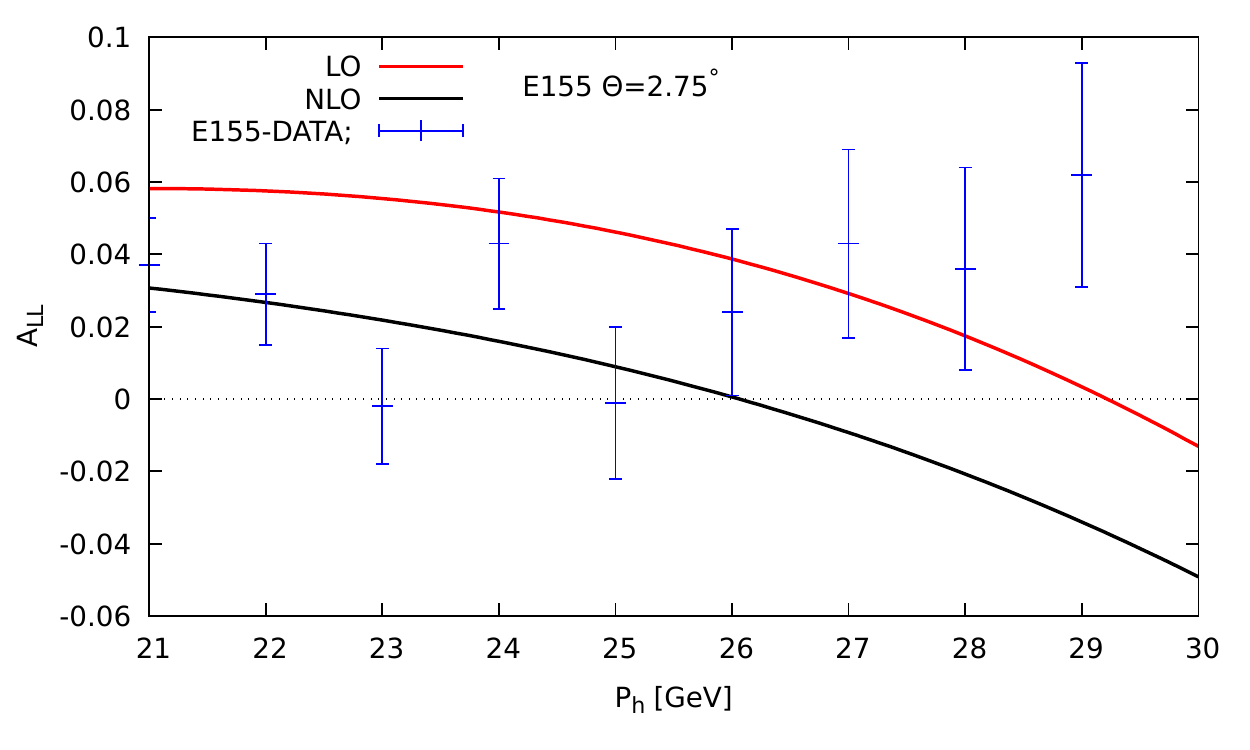}\label{fig:ALLp2pim275}}
\\
\subfloat[]{\includegraphics[width=0.5\textwidth,angle=0]{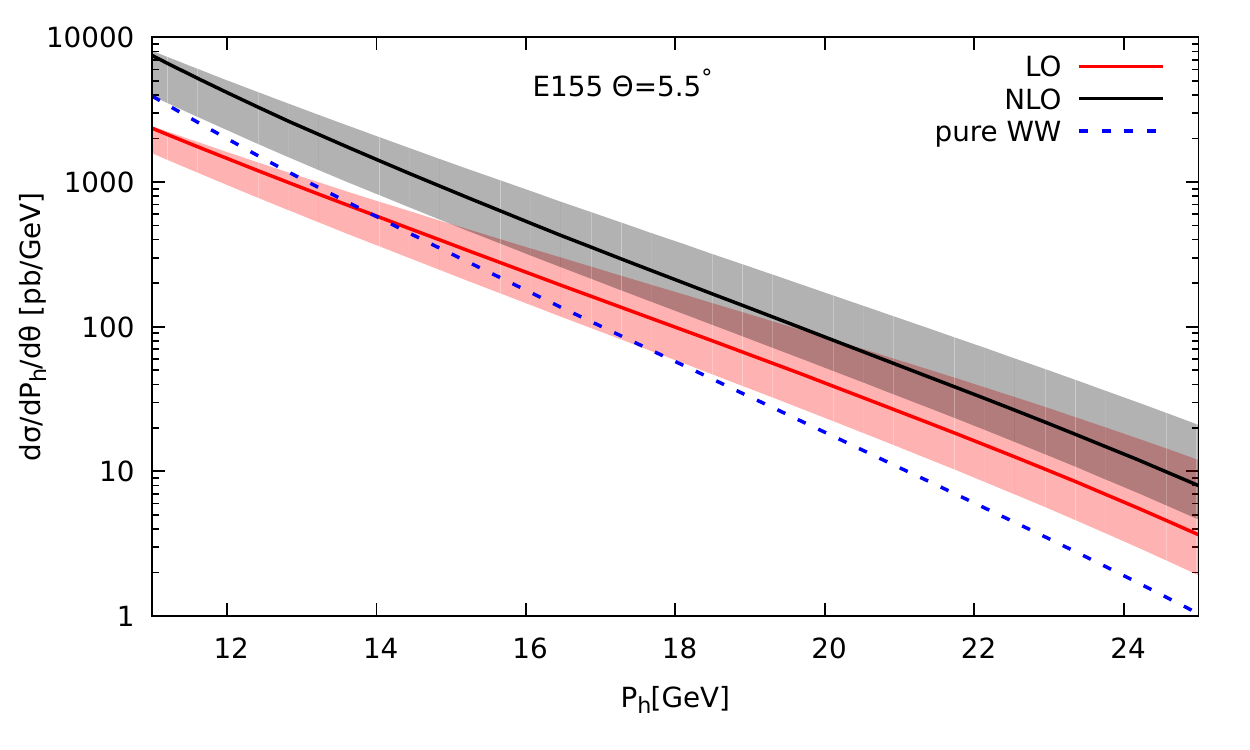}\label{fig:UUp2pim55}}
\subfloat[]{\includegraphics[width=0.5\textwidth,angle=0]{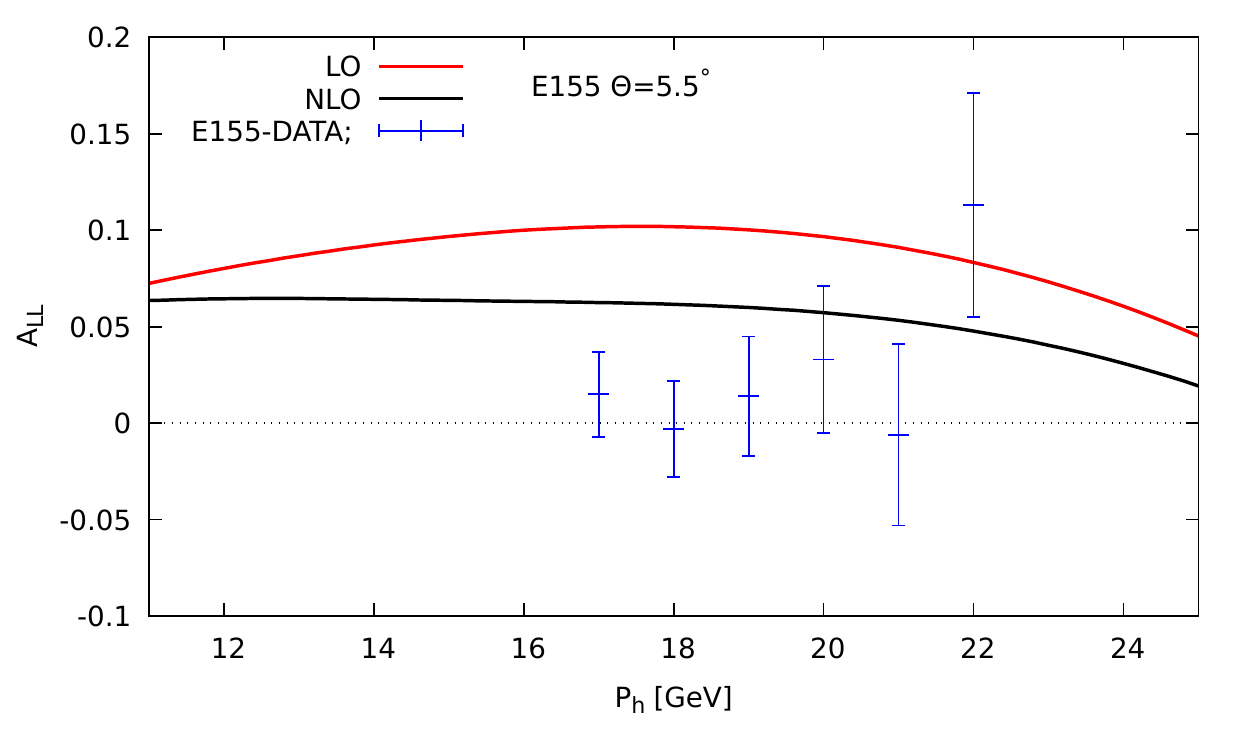}\label{fig:ALLp2pim55}}
\caption{Same as Figs.~\ref{fig:UUp2pip275} - \ref{fig:ALLp2pip55}, but for $ep\to \pi^- X$. Again, the upper panel corresponds to $\theta=2.75^{\circ}$ and the lower panel to $\theta=5.5^{\circ}$}
\end{figure*}
%%%%%%%%%%%%%%%%%%%%%%%%%%%%%%%%%%%%%%%%%%%%%%%%%%%%%%%%%%%%%

Throughout our calculations we use the NLO unpolarized parton distributions of~\cite{MSTW}, referred to as MSTW2008. 
For the helicity parton distributions we use the latest NLO set of~\cite{DSSV} (DSSV). When dealing with
deuteron targets we neglect nuclear binding effects and simply use $D=(p+n)/2$ along with the
the isospin relations $f^{u/n}=f^{d/p}$ etc. for the up and down distributions in neutrons. 
Finally, for the pion fragmentation functions we
choose the latest set of~\cite{DSSnew} (DSS14). This reference does not provide fragmentation functions for unidentified 
charged hadrons. For the latter we therefore use the earlier DSS sets~\cite{DSSold}.
%%%%%%%%%%%%%%%%%%%%%%%%%%%%%%%%%%%%%%%%%%%%%%%%%%%%%%%%%%%%%
\begin{figure*}[htb]
\centering
\subfloat[]{\includegraphics[width=0.5\textwidth,angle=0]{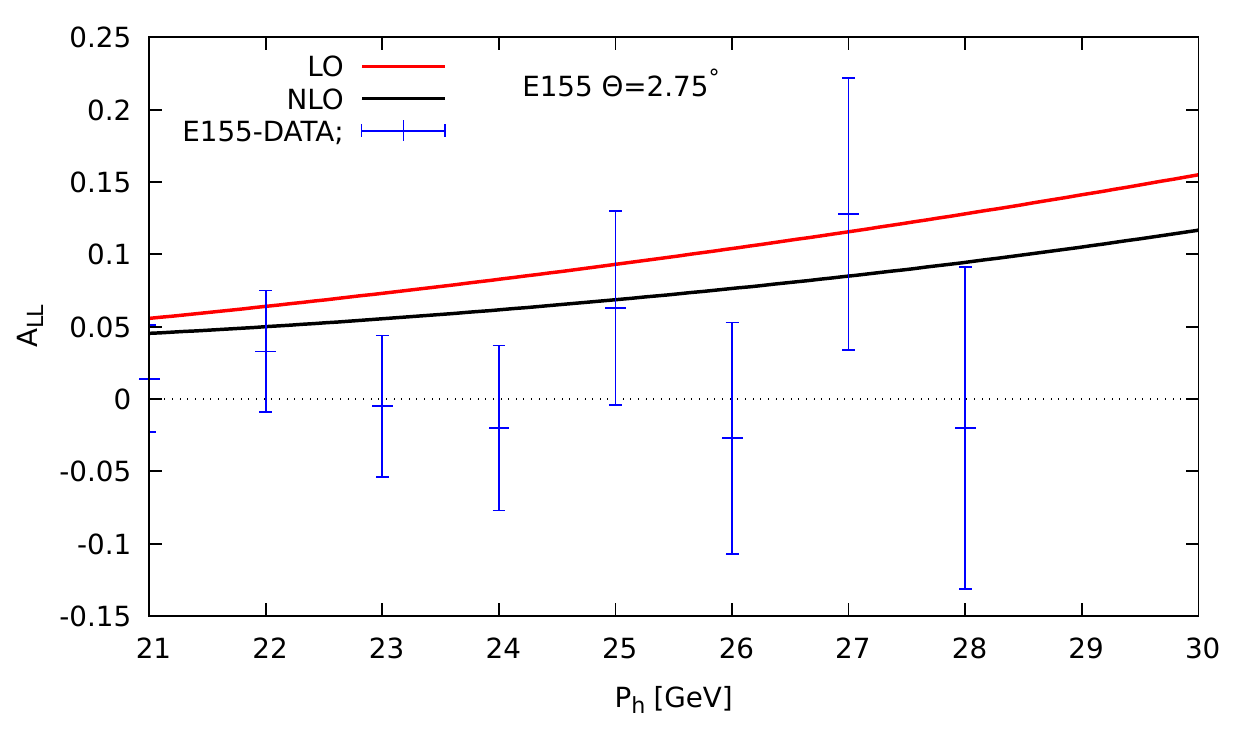}\label{fig:ALLD2pip275}}
\subfloat[]{\includegraphics[width=0.5\textwidth,angle=0]{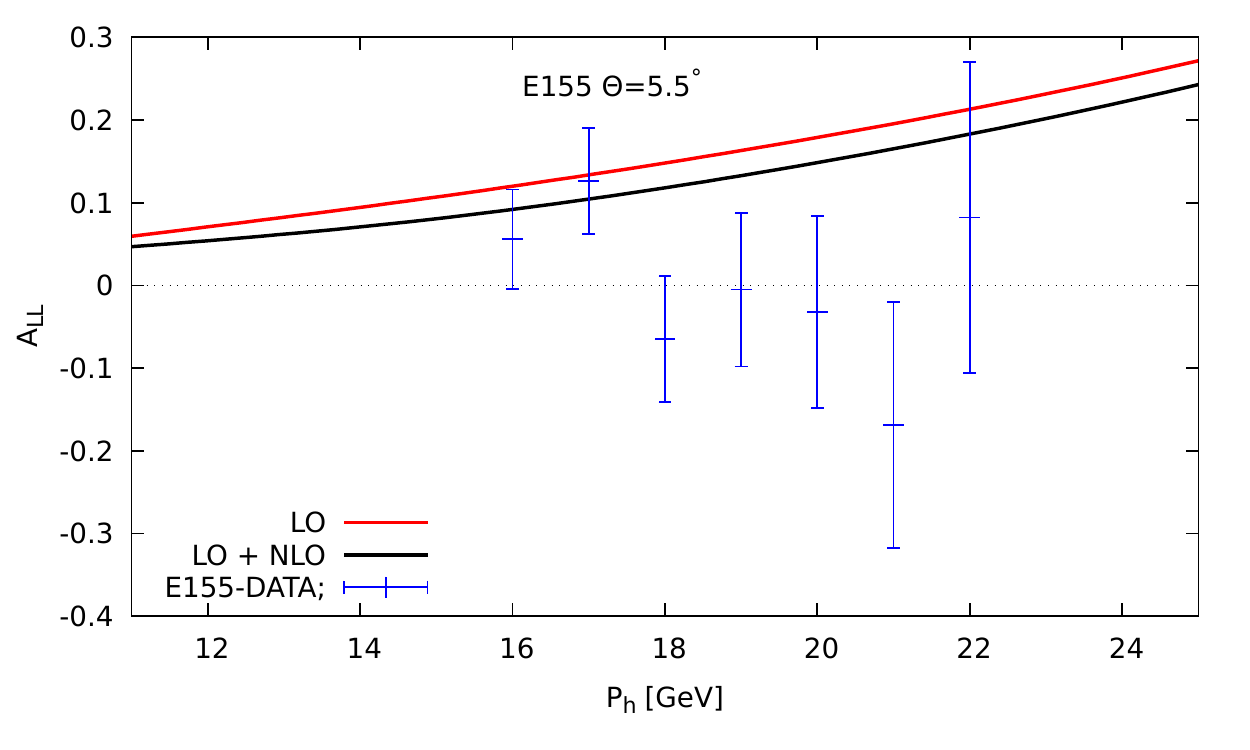}\label{fig:ALLD2pip55}}
\\
\subfloat[]{\includegraphics[width=0.5\textwidth,angle=0]{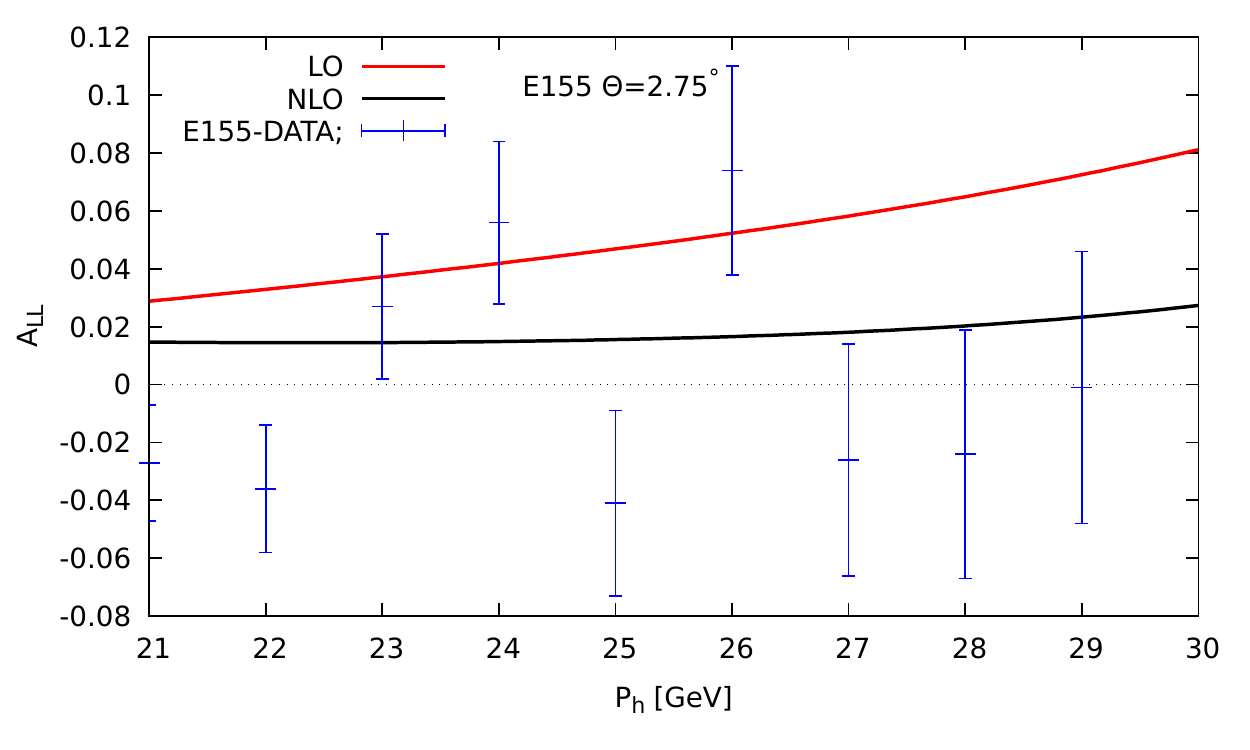}\label{fig:ALLD2pim275}}
\subfloat[]{\includegraphics[width=0.5\textwidth,angle=0]{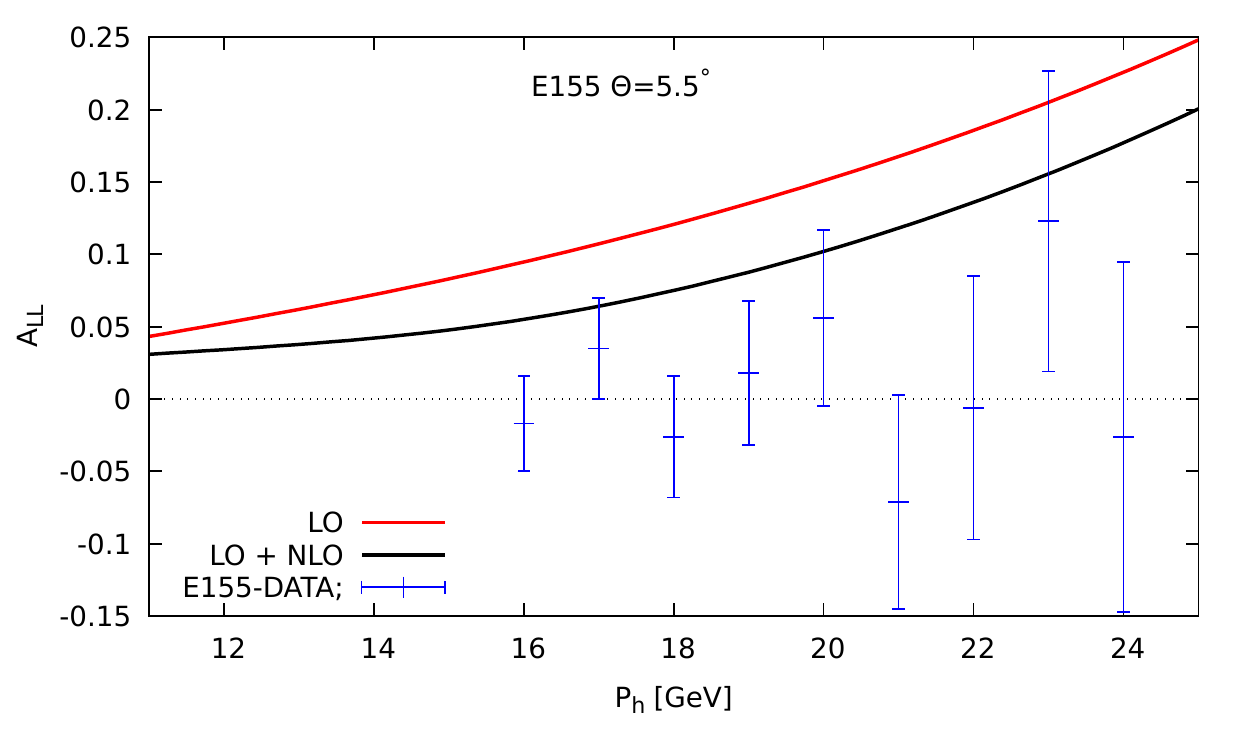}\label{fig:ALLD2pim55}}
\caption{Longitudinal double-spin asymmetries $A_{\mathrm{LL}}$ for $eD\to\pi^{+}X$ (a),(b) and $eD\to\pi^{-}X$ (c),(d) for the scattering angles $\theta=2.75^{\circ}$ (a),(c) and $\theta=5.5^{\circ}$ (b),(d).}
\end{figure*}
%%%%%%%%%%%%%%%%%%%%%%%%%%%%%%%%%%%%%%%%%%%%%%%%%%%%%%%%%%%%%

The experiment E155 has released data for the channels $ep\to \pi^{\pm}X$, $ep\to h^{\pm}X$, 
$eD\to \pi^{\pm}X$ and $eD\to h^{\pm}X$. In the following we will briefly discuss each of these.

\subsubsection{$ep\to \pi^+X$}

Figure~\ref{fig:UUp2pip275} shows the unpolarized cross section for $ep\to \pi^+X$ as a function of
$|\vec{P}_{h\perp}|$ at a scattering angle of $\theta=2.75^\circ$. We plot
\begin{equation}
\frac{d\sigma^{\ell N\to hX}}{d|\vec{P}_h|\,d\theta} = 2\pi |\vec{P}_h|\sin(\theta)\,
\left(E_h\frac{d\sigma^{\ell N\to hX}}{d^3 \vec{P}_h}\right),\label{CSPthE155}
\end{equation}
at LO (lower solid line and band) and NLO  (upper). The solid lines represent the cross sections computed at 
scale $\mu=|\vec{P}_{h\perp}|=|\vec{P}_h|\sin\theta$, while the bands are generated from the scale 
variation $1\,\mathrm{GeV}<\mu<2 P_{h\perp}$. The upper end of the band corresponds to the lowest scale. 
We find that the NLO corrections are large, with $K\equiv \sigma_{\mathrm{NLO}}/\sigma_{\mathrm{LO}}$-factors 
of about 2-3 at E155 for this scattering angle. The dashed line in Fig.~\ref{fig:UUp2pip275} 
represents the contribution to the cross section by quasi-real photons, obtained in the unpolarized
case as (cf. Eq.~(\ref{Trafox1}) and see also Eq.~(25) of Ref.~\cite{ourpaper}):
\beeq 
\sigma&=&\left(\frac{-U}{S^2}\right)\sum_{i, f}
\int_{\frac{U}{T+U}}^{1+\frac{T}{S}} \frac{dv}{v(1-v)}\int_{\tfrac{1-v}{v}\tfrac{U}{T}}^1 \frac{dw}{w^2} \,
\frac{f^{i/N}(x,\mu)}{x}\nonumber\\[2mm]
& \times &\frac{D^{h/f}(z,\mu)}{z^2} \,f^{\gamma/\ell}\left(\tfrac{1-v}{1-vw},\mu\right)\,
\frac{\alpha_s(\mu)}{\pi}\,\hat{\sigma}_{\mathrm{LO}}^{\gamma i\to f}(v,w)\,.
\label{Trafox1a}
\eeeq
We refer to this contribution in the plots as ``pure WW'' contribution. We again adopt the scale
$\mu=|\vec{P}_{h\perp}|$. As one can see, the real-photon contribution dominates the
cross section only at the lower values of $|\vec{P}_h|$. 
This finding is at variance with the general assumption made in~\cite{E155,Afanasev:1996mj,ep7} that the
bulk of inclusive-hadron events is produced by real photons. 
%We note that for scattering angle $\theta=2.75^\circ$
%real photons do dominate in the region where most of the data were taken. 
%This becomes different 
For the scattering angle $\theta=5.5^\circ$ (see Fig.~\ref{fig:UUp2pip55})
the contribution by quasi-real photons does not really dominate anywhere in the
regime of interest. We conclude that our full NLO calculation is required here for a meaningful 
comparison to the data. We note that the $K$-factors are slightly smaller at this scattering
angle. 
%%%%%%%%%%%%%%%%%%%%%%%%%%%%%%%%%%%%%%%%%%%%%%%%%%%%%%%%%%%%%
\begin{figure*}[htb]
\centering
\subfloat[]{\includegraphics[width=0.5\textwidth,angle=0]{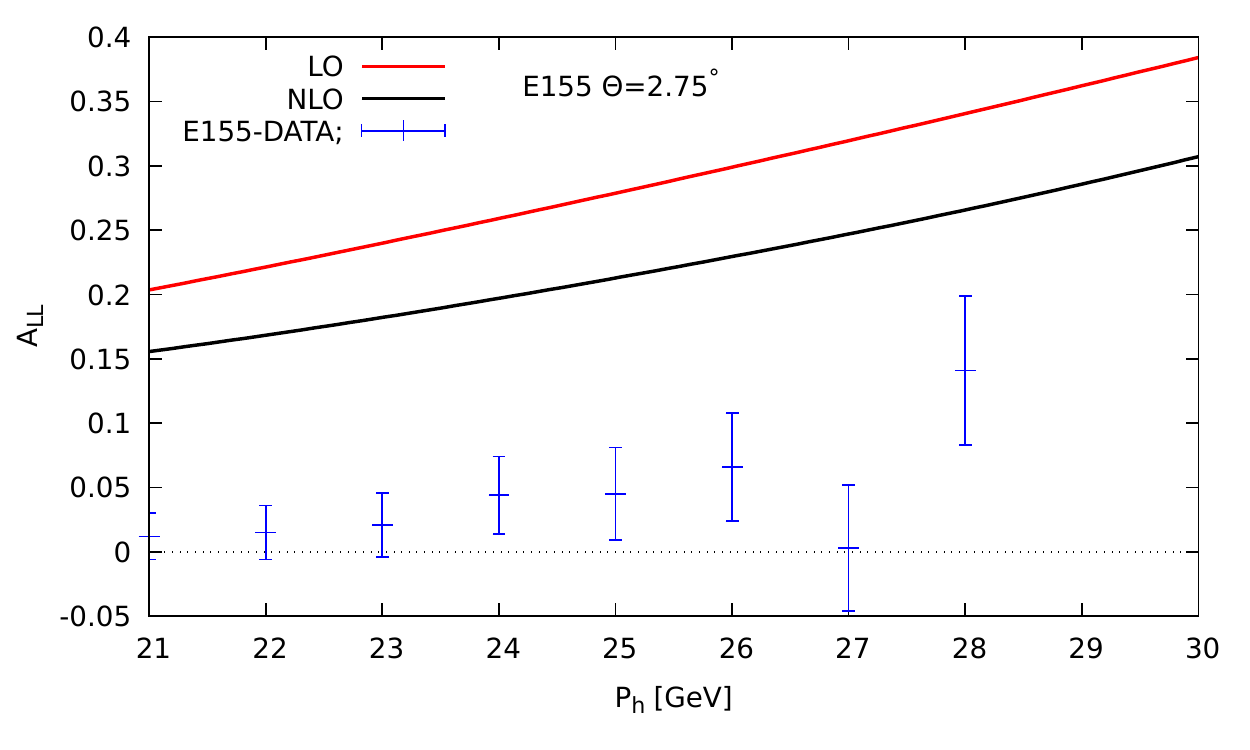}\label{fig:ALLp2hp275}}
\subfloat[]{\includegraphics[width=0.5\textwidth,angle=0]{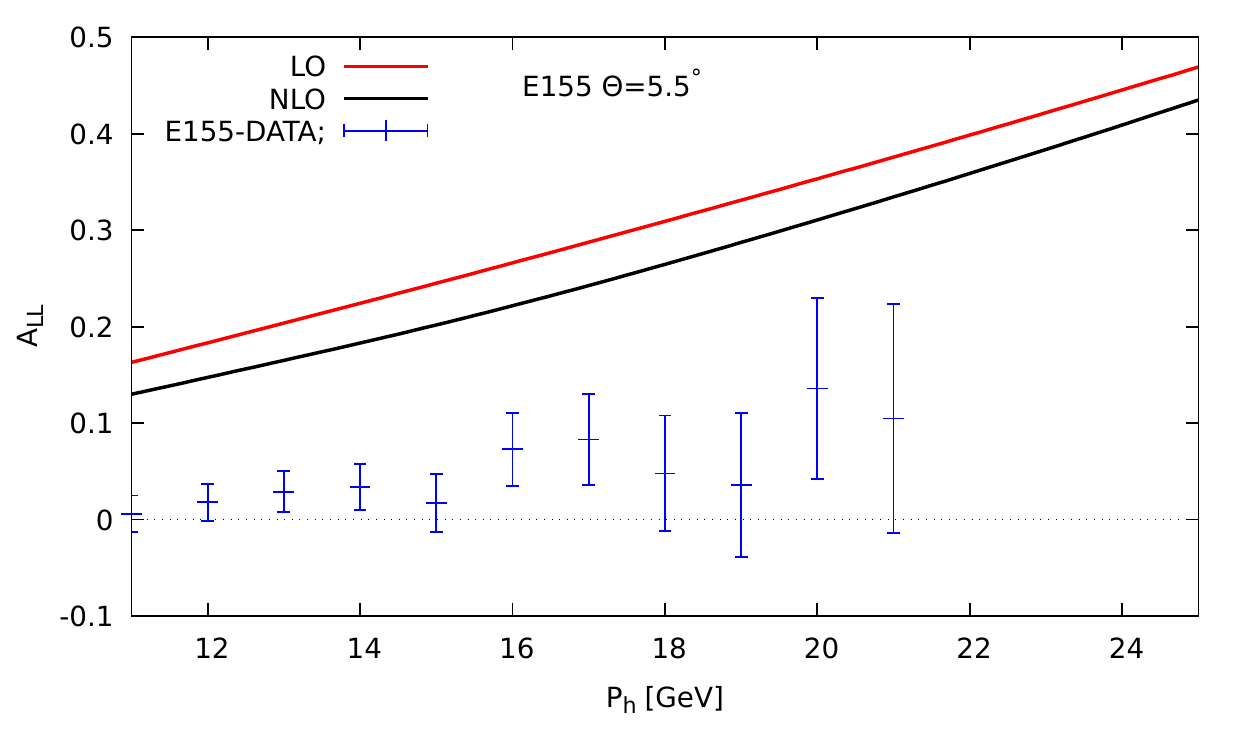}\label{fig:ALLp2hp55}}
\\
\subfloat[]{\includegraphics[width=0.5\textwidth,angle=0]{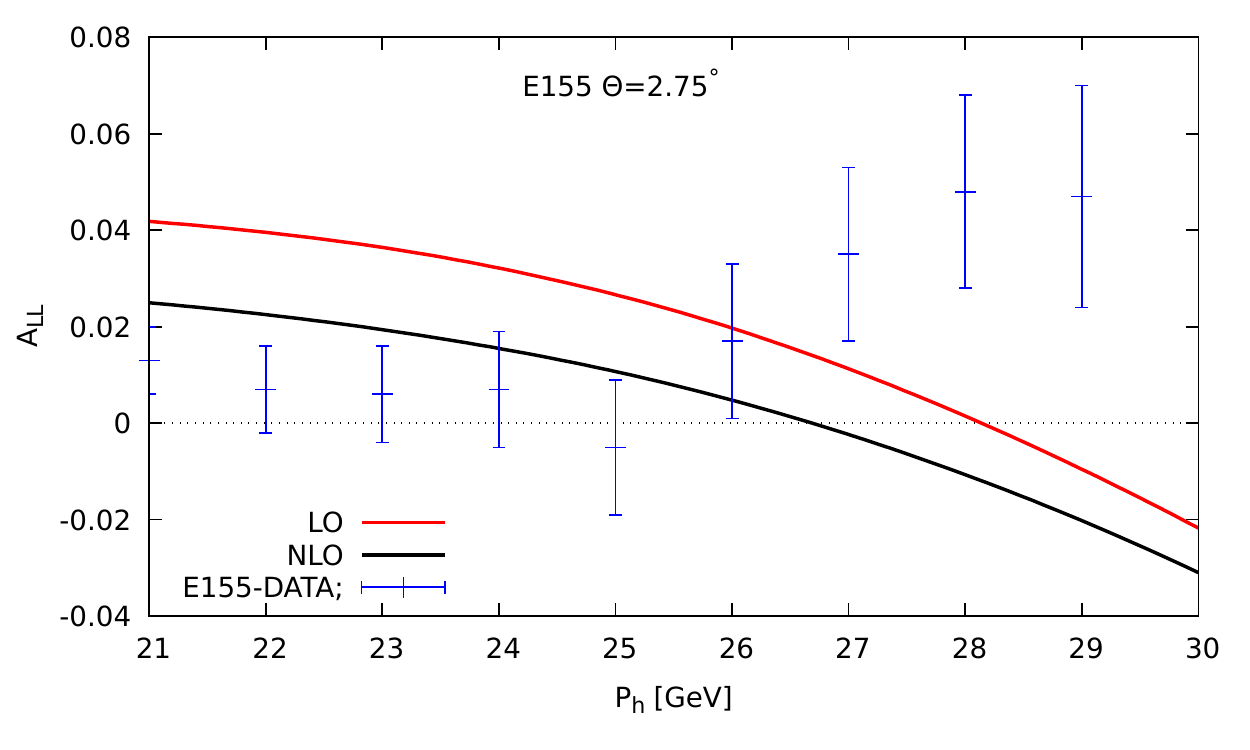}\label{fig:ALLp2hm275}}
\subfloat[]{\includegraphics[width=0.5\textwidth,angle=0]{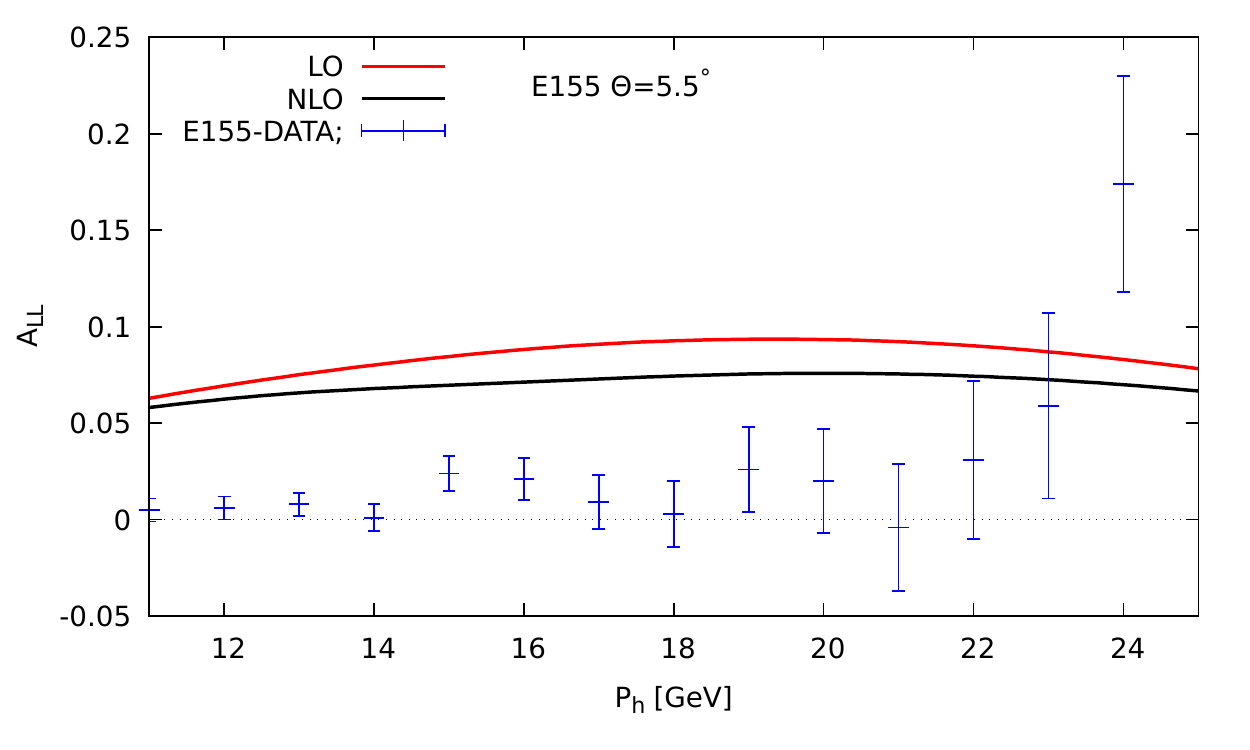}\label{fig:ALLp2hm55}}
\caption{Same as Figs.~\ref{fig:ALLD2pip275} - \ref{fig:ALLD2pim55}, but for production of 
unidentified positive (a),(b) and negative (c),(d) hadrons off a proton target.}
\end{figure*}
%%%%%%%%%%%%%%%%%%%%%%%%%%%%%%%%%%%%%%%%%%%%%%%%%%%%%%%%%%%%%

Turning to the corresponding spin asymmetries shown in Figs.~\ref{fig:ALLp2pip275}, \ref{fig:ALLp2pip55} we find that the 
NLO corrections do not influence the asymmetries as much as the cross sections. Instead, a significant part of 
NLO corrections seems to cancel in the asymmetry. On the other hand, there is a clear trend for
the asymmetry to decrease when going from LO to NLO. This helps to bring the theoretical results
closer to the data. Still, even at NLO our results for the spin asymmetry are much higher than the
data for $\theta=2.75^\circ$. For the angle $\theta=5.5^\circ$ we find a slightly better agreement, mostly
because the data have larger error bars here. We note that for the kinematics that are relevant here
the involved parton distributions and fragmentation functions are rather well constrained. It is 
conceivable that the disagreement we observe for the spin asymmetries indicates that perturbative-QCD
methods are not yet applicable at such relatively low $|\vec{P}_{h\perp}|$. Higher-twist 
corrections might account for the difference, in particular for the data at $\theta=2.75^\circ$.

%%%%%%%%%%%%%%%%%%%%%%%%%%%%%%%%%%%%%%%%%%%%%%%%%%%%%%%%%%%%%
\begin{figure*}[htb]
\centering
\subfloat[]{\includegraphics[width=0.5\textwidth,angle=0]{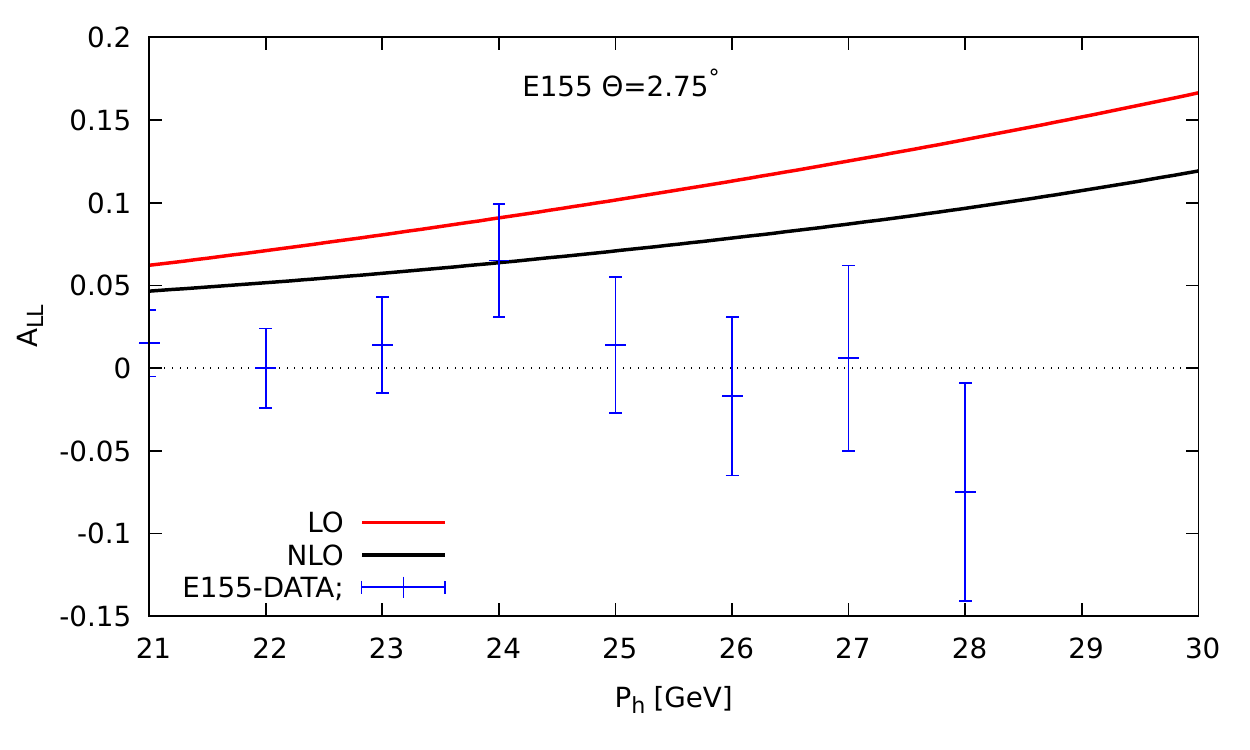}\label{fig:ALLD2hp275}}
\subfloat[]{\includegraphics[width=0.5\textwidth,angle=0]{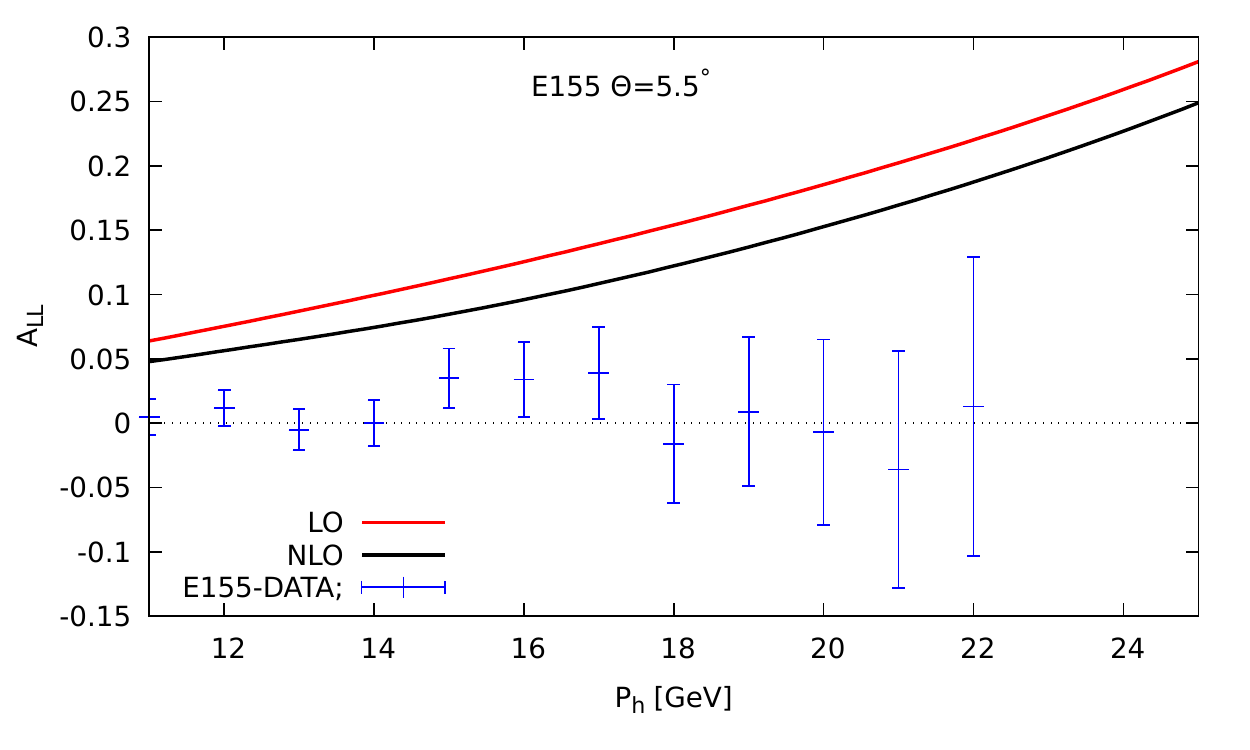}\label{fig:ALLD2hp55}}
\\
\subfloat[]{\includegraphics[width=0.5\textwidth,angle=0]{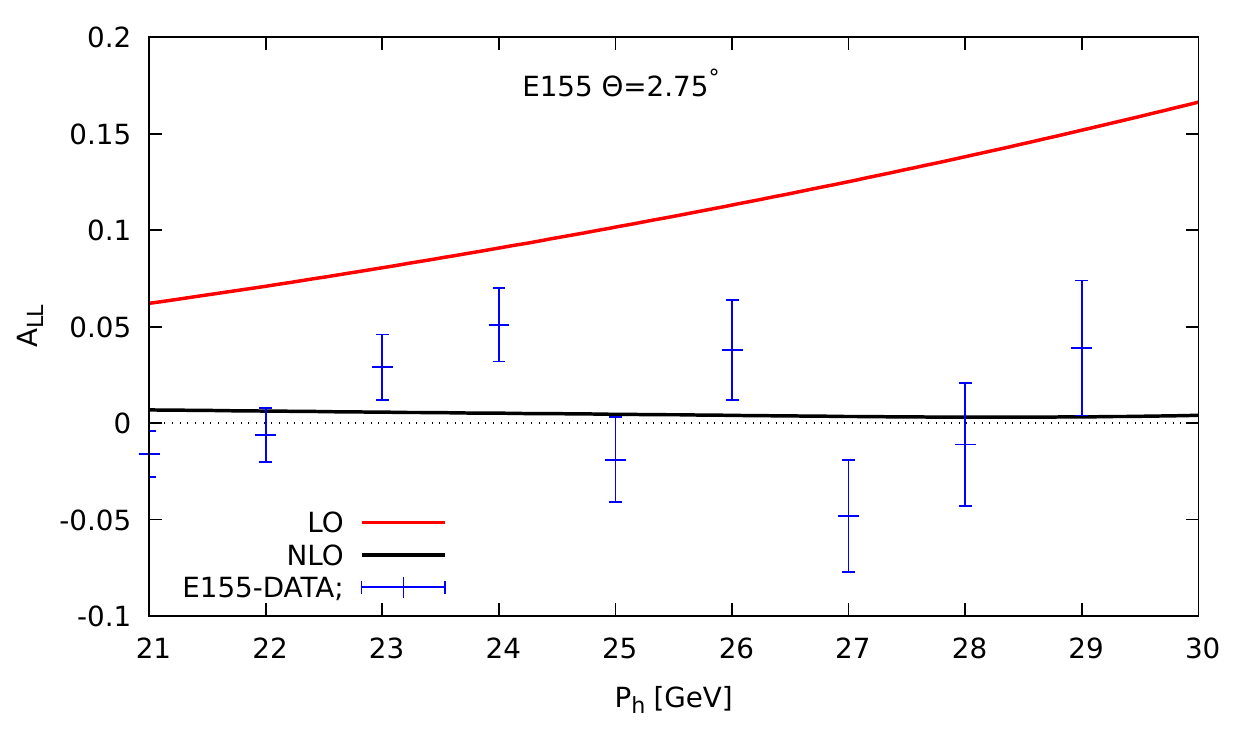}\label{fig:ALLD2hm275}}
\subfloat[]{\includegraphics[width=0.5\textwidth,angle=0]{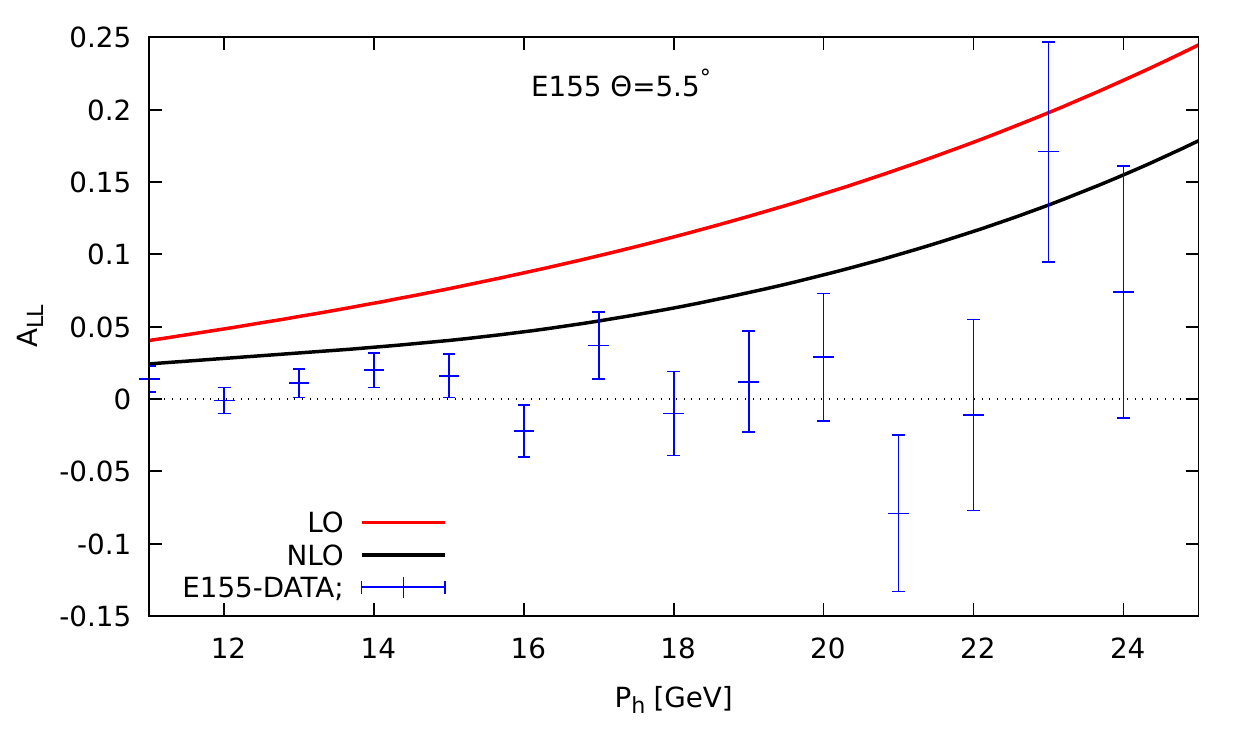}\label{fig:ALLD2hm55}}
\caption{Same as Figs.~\ref{fig:ALLD2pip275} - \ref{fig:ALLD2pim55}, but for unidentified positive (a), (b) and 
negative (c), (d) hadrons produced off a deuteron target.}
\end{figure*}
%%%%%%%%%%%%%%%%%%%%%%%%%%%%%%%%%%%%%%%%%%%%%%%%%%%%%%%%%%%%%

\subsubsection{$ep\to \pi^-X$}

In Figs.~\ref{fig:UUp2pim275} -- \ref{fig:ALLp2pim55} we present our results for $\pi^-$ production off 
a proton target. The plots of the unpolarized cross sections in Figs.~\ref{fig:UUp2pim275}, \ref{fig:UUp2pim55} 
qualitatively resemble those for $\pi^+$-production, that is, we observe large $K$-factors and dominance of the
real-photon contribution at the smaller $\pi^-$ momenta for $\theta=2.75^\circ$. Also, as for $\pi^+$-production 
the contributions by real photons do not dominate for $\theta=5.5^\circ$. Since we find the same qualitative 
features of the cross section also for all other channels, $eD\to\pi^{\pm}X$, $ep\to h^{\pm}X$ and $eD\to h^{\pm}X$,
we refrain from showing plots for their unpolarized cross sections.

%%%%%%%%%%%%%%%%%%%%%%%%%%%%%%%%%%%%%%%%%%%%%%%%%%%%%%%%%%%%
\begin{figure*}[htb]
\centering
\hspace*{-0.9cm}
\subfloat[]{\includegraphics[width=0.5\textwidth,angle=0]{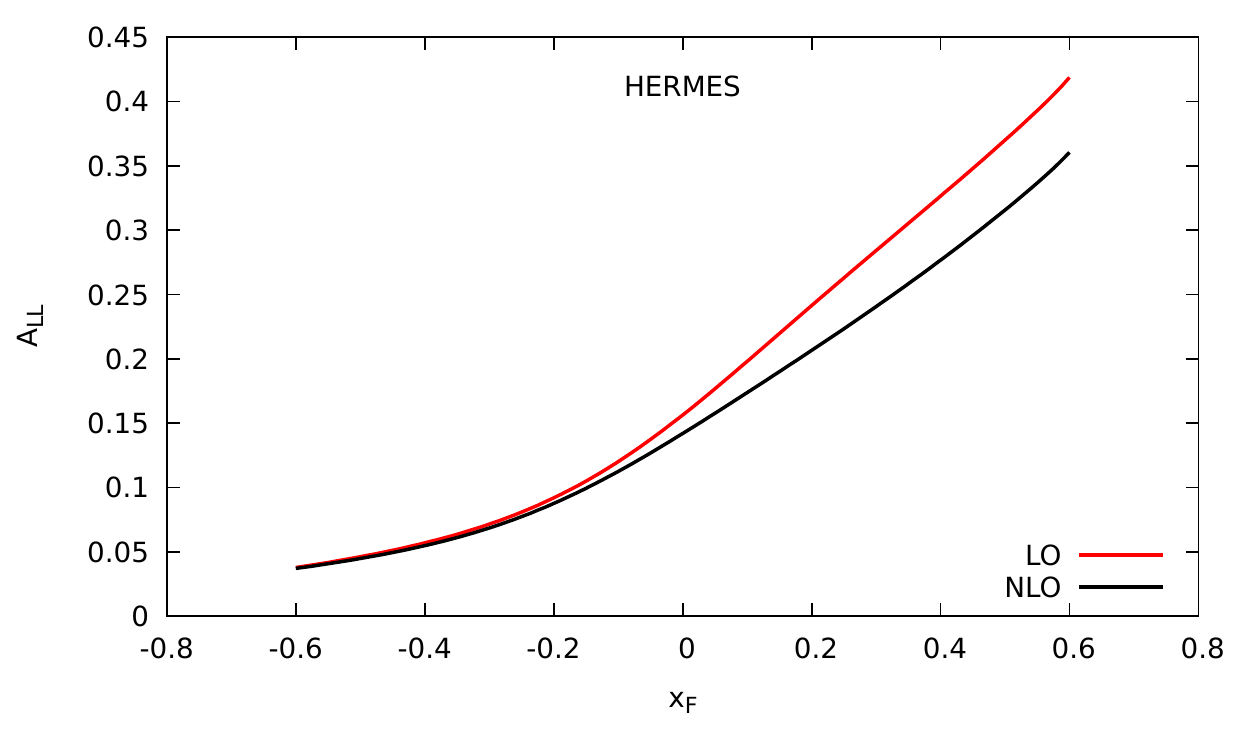}\label{fig:HERMESxf}}
\hspace*{0.0\textwidth}
\subfloat[]{\includegraphics[width=0.5\textwidth,angle=0]{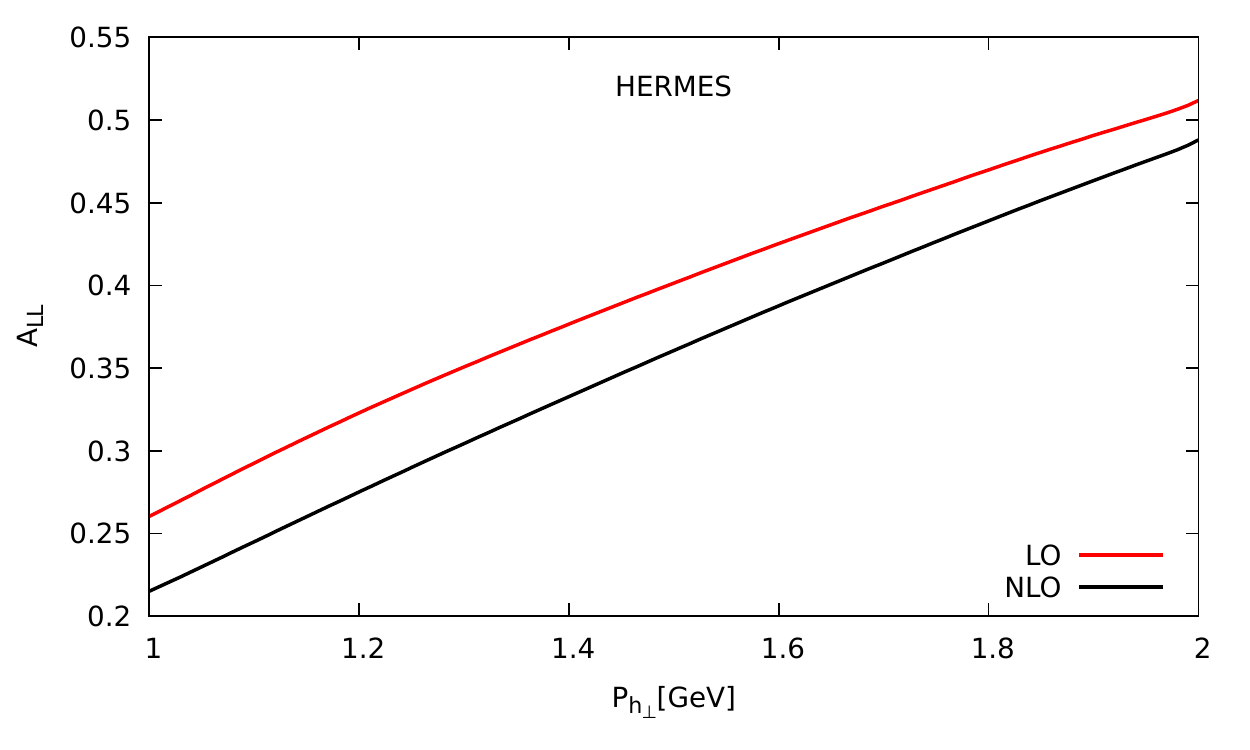}\label{fig:HERMESpt}}
\caption{Longitudinal double-spin asymmetries $A_{\mathrm{LL}}$ for $e p\to \pi^+X$ at HERMES, at LO and NLO, 
(a) as function of $x_F$ for 
$1\;\mathrm{GeV}<P_{h\perp}<2.2\;\mathrm{GeV}$, and (b) as function of $P_{h\perp}$ for $0.3<x_F<0.55$.}
\end{figure*}
%%%%%%%%%%%%%%%%%%%%%%%%%%%%%%%%%%%%%%%%%%%%%%%%%%%%%%%%%%%%

In Figs.~\ref{fig:ALLp2pim275}, \ref{fig:ALLp2pim55} we compare our results for the asymmetries to the 
E155 data and find a better agreement with the data than for $\pi^+$-production. Again we observe that the 
NLO corrections are overall not quite as large for the asymmetries as they are for the cross sections. As before they tend to 
push the theory curves closer to the data. 

\subsubsection{$eD\to \pi^{\pm}X$}

Figures~\ref{fig:ALLD2pip275} -- \ref{fig:ALLD2pim55} present numerical results for the asymmetries 
$A_{\mathrm{LL}}$ for pion production off a deuteron target. We observe an overall better agreement
with the E155 data than for scattering off protons, especially for $\pi^+$-production. Again, the NLO
corrections tend to improve the agreement, although by and large, the NLO results are somewhat higher than 
the data.

\subsubsection{$ep\to h^{\pm}X$}

We finally discuss the spin asymmetries for unidentified charged hadrons. Our results for production off a proton 
target are shown in Figs.~\ref{fig:ALLp2hp275} -- \ref{fig:ALLp2hm55}. We find that for positively charged hadrons 
the NLO results are much higher than the E155 data for both scattering angles $\theta=2.75^\circ$ and $\theta=5.5^\circ$.
This is also true for negatively charged hadrons at the larger scattering angle. For $ep\to h^-X$ at $\theta=2.75^\circ$
the trend of the data as a function of the hadron momentum is opposite to that of the theoretical results.

\subsubsection{$eD\to h^{\pm}X$}

The corresponding results for unidentified hadrons produced off a deuteron target are shown 
in Figs.~\ref{fig:ALLD2hp275} -- \ref{fig:ALLD2hm55}. Compared to the case of a proton target
the theoretical curves are now overall much closer to the E155 data. This finding is in line
with what we observed for pion production above. 

\section{Predictions \label{pheno}}

In view of the unclear situation concerning the comparison of NLO theory and E155 data we 
argue that it would be important to have independent data on the unpolarized cross section and the longitudinal double-spin 
asymmetry.  We now present some phenomenological predictions for $A_{\mathrm{LL}}$ at NLO in single-inclusive pion 
production for HERMES, JLab12, COMPASS and the future EIC. For the latter, we also investigate the spin asymmetry in jet production.
In our previous paper~\cite{ourpaper} we have already presented results for the corresponding spin-averaged cross sections,
and we compute the spin asymmetry for the same kinematics considered there. We always show LO and full NLO
results, using the scale $\mu=|\vec{P}_{h\perp}|$.

%%%%%%%%%%%%%%%%%%%%%%%%%%%%%%%%%%%%%%%%%%%%%%%%%%%%%%%%%%%%%
\begin{figure*}[htb]
\centering
\hspace*{-0.9cm}
\subfloat[]{\includegraphics[width=0.5\textwidth,angle=0]{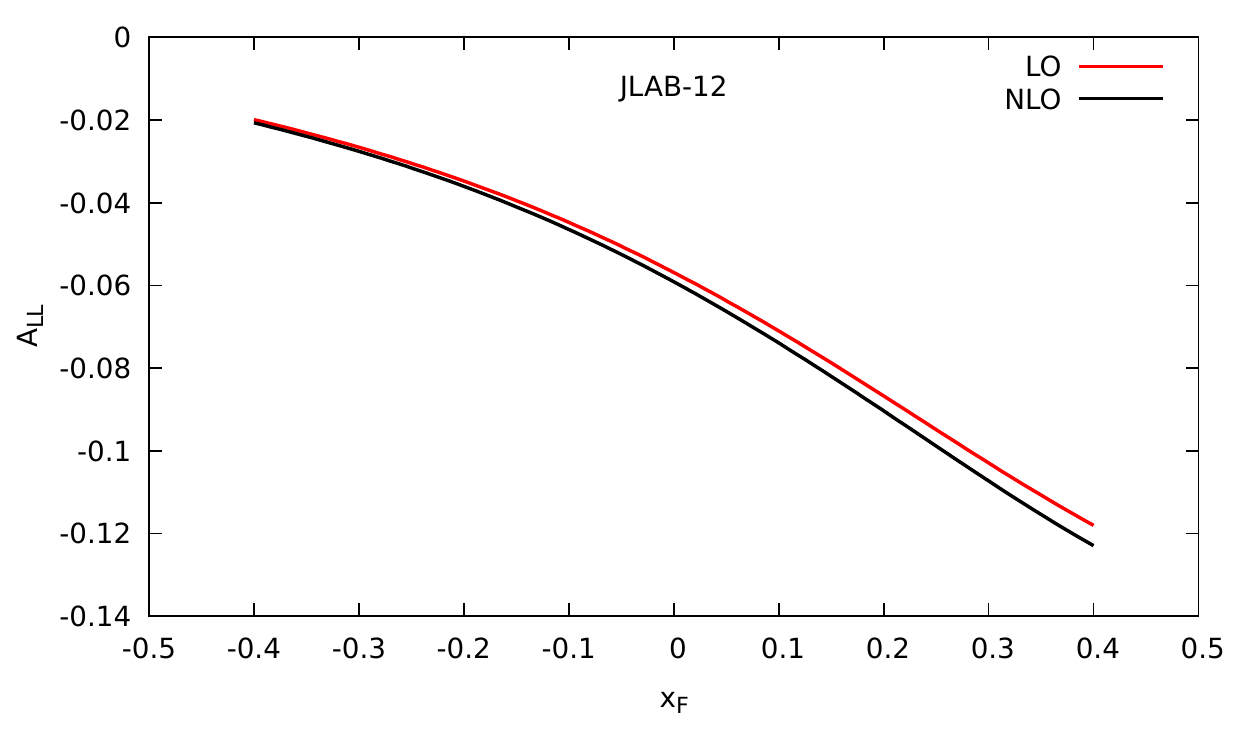}\label{fig:JLabxf}}
\subfloat[]{\includegraphics[width=0.5\textwidth,angle=0]{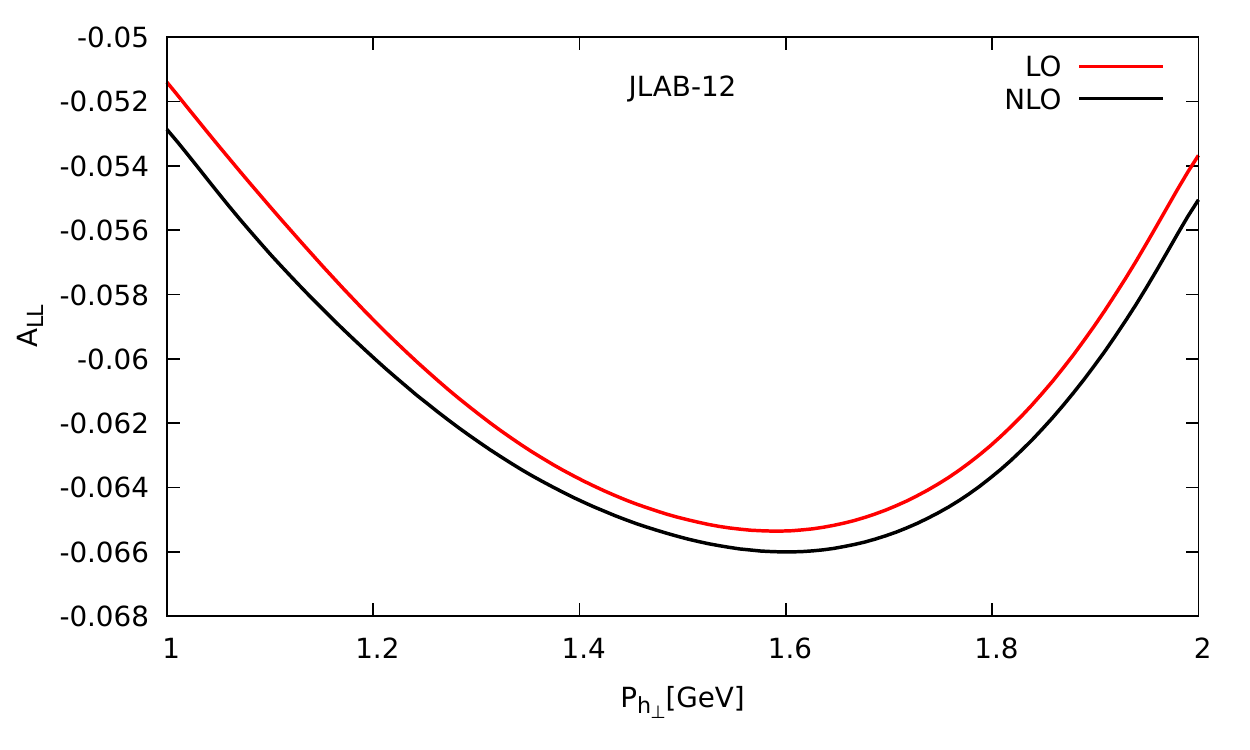}\label{fig:JLabpt}}
\caption{Same as Figs.~\ref{fig:HERMESxf},~\ref{fig:HERMESpt}, but for $\ell \,{}^3{\mathrm{He}}$ scattering
at beam energy 12 GeV after the CEBAF upgrade at Jefferson Lab. On the left we have chosen 
a fixed $P_{h\perp}=1.5\; \mathrm{GeV}$, while for the $P_{h\perp}$ dependence on the right we have
integrated over $-0.4\leq x_F\leq 0.4$.}
\end{figure*}
%%%%%%%%%%%%%%%%%%%%%%%%%%%%%%%%%%%%%%%%%%%%%%%%%%%%%%%%%%%%%

In Figs.~\ref{fig:HERMESxf}, \ref{fig:HERMESpt} we show the asymmetries in $ep\to \pi^+ X$ at
$\sqrt{S}=7.25$~GeV, as relevant for HERMES. The left figure shows the dependence of $A_{\mathrm{LL}}$ on the Feynman 
variable $x_F$, averaging the cross sections over $1\,\mathrm{GeV}<P_{h\perp}<2.2\,\mathrm{GeV}$.
Similar to what we found for E155 the NLO corrections to the asymmetry are not large, despite large $K$-factors 
for the spin-averaged cross section (cf. Ref.~\cite{ourpaper}). We observe that the asymmetries grow toward
larger Feynman-$x_F$ where the NLO corrections become larger. Figure~\ref{fig:HERMESpt} shows the 
$P_{h\perp}$-dependence of $A_{\mathrm{LL}}$,  with $x_F$ averaged over $0.3<x_F<0.55$.
Clearly, a very large spin asymmetry is expected for these kinematics. 

Figures~\ref{fig:JLabxf}, \ref{fig:JLabpt} show results for the spin asymmetry in $\ell\,{}^3\mathrm{He}$
scattering at beam energy of 12 GeV, corresponding to measurements possible with the CEBAF upgrade at Jefferson Lab.
For the calculations of the unpolarized cross section we neglect nuclear effects for Helium and just set
${}^3$He$=(2p+n)/3$ along with the usual isospin relations for the parton distributions. 
The situation is different for the helicity distributions. To a good approximation the two spins of the protons in a
polarized ${}^3$He nucleus are antiparallel. Effectively, the nucleus can be
considered a polarized neutron target, with $\Delta f^{q/^3\mathrm{He}}=\Delta f^{q/n}$. 
Again, we then use isospin symmetry to obtain the neutron's helicity distributions.
As seen from Figs.~\ref{fig:JLabxf}, \ref{fig:JLabpt}, the resulting asymmetry is negative and 
much smaller in size than the one for a proton target found for HERMES kinematics. 
The NLO corrections affect the asymmetry only little. We stress that for the very modest beam
energy at JLab12 the use of perturbative methods for analyzing the process $\ell N\to hX$
is questionable.

%%%%%%%%%%%%%%%%%%%%%%%%%%%%%%%%%%%%%%%%%%%%%%%%%%%%%%%%%
\begin{figure*}[htb]
\centering
\hspace*{-0.9cm}
\subfloat[]{\includegraphics[width=0.5\textwidth,angle=0]{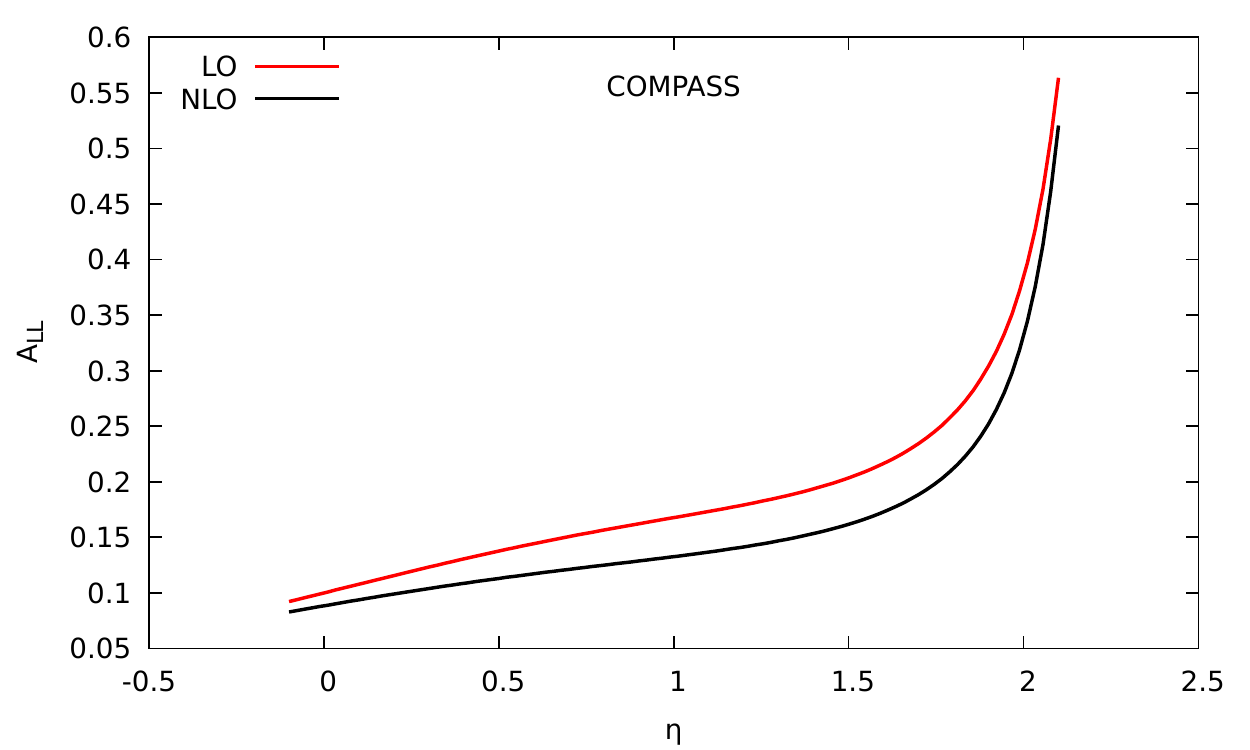}\label{fig:COMPASSeta}}
\subfloat[]{\includegraphics[width=0.5\textwidth,angle=0]{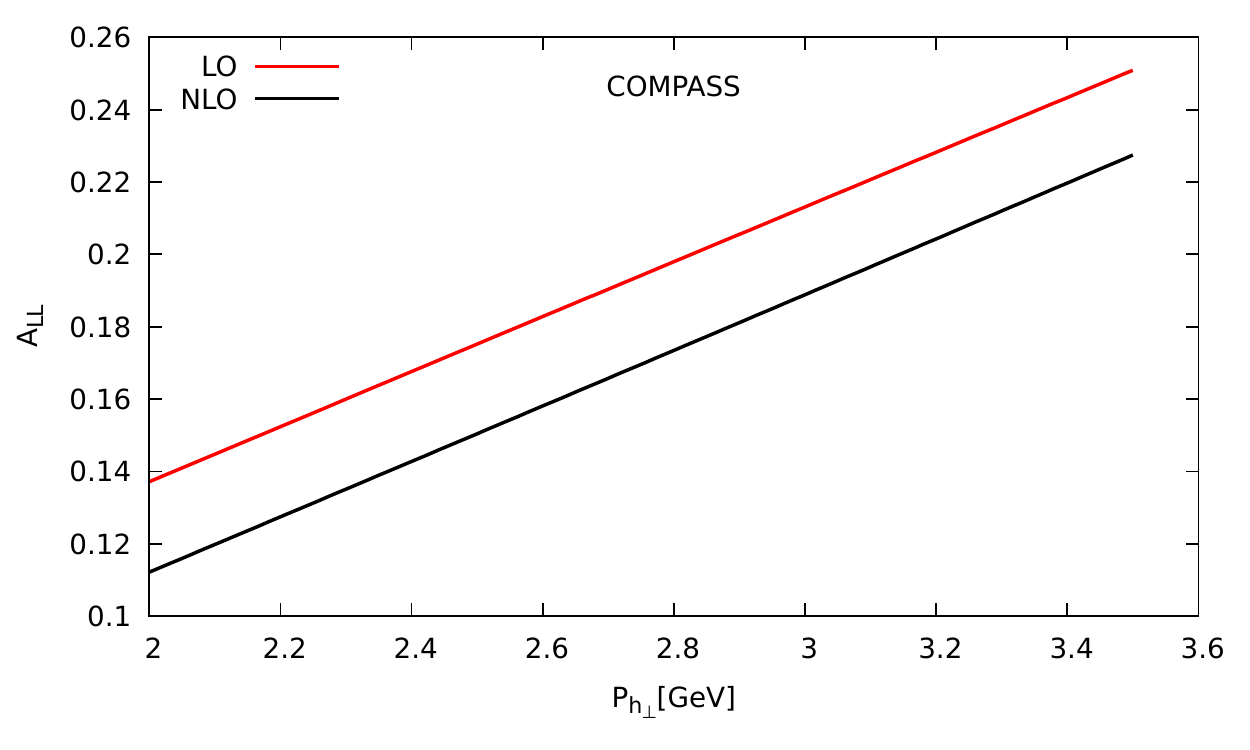}\label{fig:COMPASSpt}}
\caption{Longitudinal double-spin asymmetry $A_{\mathrm{LL}}$ for $\mu p\to \pi^0X$ at COMPASS, at LO
and NLO, (a) as function of pion pseudorapidity for fixed 
$P_{h\perp}=2\;\mathrm{GeV}$, and (b) as function of $P_{h\perp}$ for $-0.1\leq \eta\leq 2.38$.}
\end{figure*}
%%%%%%%%%%%%%%%%%%%%%%%%%%%%%%%%%%%%%%%%%%%%%%%%%%%%%%%%%

The COMPASS experiment employs  a $160\,\mathrm{GeV}$ muon beam on a fixed target, resulting in 
a much larger center-of-mass energy of $\sqrt{S}=17.4\,\mathrm{GeV}$. As a result, a wider $P_{h\perp}$-range can be probed. 
As we found in Ref.~\cite{ourpaper}, this yields a more controlled perturbative framework, with the NLO corrections
to the spin-averaged cross section amounting to only about $30-40\%$. Predictions for $A_{\mathrm{LL}}$ 
in $\mu p\to \pi^0 X$ at COMPASS are shown in Figs.~\ref{fig:COMPASSeta}, \ref{fig:COMPASSpt}. On the left we plot 
the asymmetry as a function of pion's c.m.s. pseudorapidity $\eta$, at a fixed transverse momentum $P_{h\perp}=2\,\mathrm{GeV}$. 
On the right, we show the $P_{h\perp}$-dependence of $A_{\mathrm{LL}}$, averaging over $-0.1\leq \eta\leq 2.38$.
We find that the asymmetry is again reduced by the NLO corrections. Despite the relatively large energy
$A_{\mathrm{LL}}$ is expected to be sizable. 

Excellent opportunities for studies of single-inclusive hadron production would be provided by a 
future EIC~\cite{Accardi:2012qut}. Thanks to the high $ep$ c.m.s. energy of an EIC, $\sqrt{S}=100$~GeV, 
it will become possible to probe much larger transverse hadron momenta where pQCD is expected to work better. We have presented numerical results in Ref.~\cite{ourpaper} (see Fig.~8a of this reference) for the $\eta$-dependence of the unpolarized cross section for $e p\to \pi^+X$ at the EIC, at a relatively large fixed transverse hadron momentum $P_{h\perp}=10\,\mathrm{GeV}$. Our results indicated a milder modification of the LO result by NLO corrections for such a large transverse hadron momentum, with a $K$-factor of about 1.5. In particular, the NLO corrections are dominated by real photon contributions for positive pseudorapidities $\eta>1$. We find for the $\eta$-dependence of the asymmetry $A_{LL}$ at $P_{h\perp}=10\,\mathrm{GeV}$ that the effect of the NLO corrections is rather small.

\begin{figure*}[htb]
\centering
\hspace*{-0.9cm}
\subfloat[]{\includegraphics[width=0.5\textwidth,angle=0]{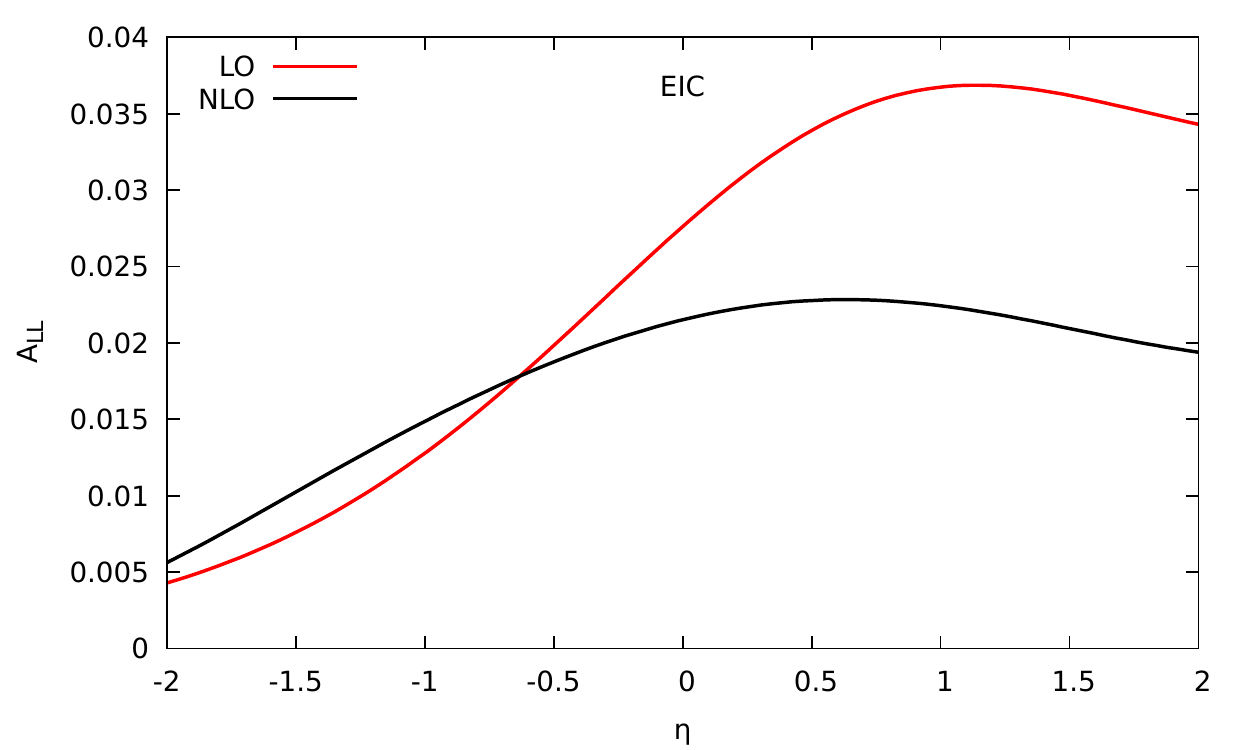}\label{fig:EICeta}}
\subfloat[]{\includegraphics[width=0.5\textwidth,angle=0]{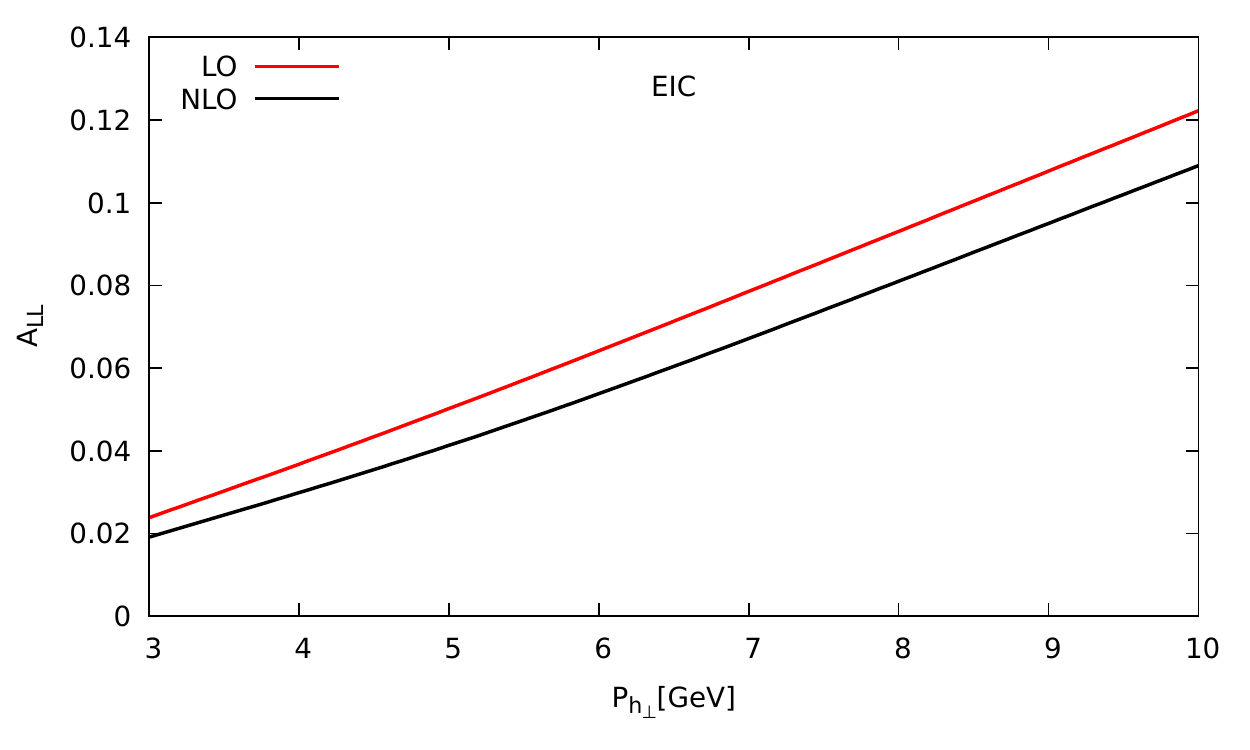}\label{fig:EICpt}}
\caption{Longitudinal double-spin asymmetry $A_{\mathrm{LL}}$ for $e p\to \pi^+X$ at an EIC with 
$\sqrt{S}=100$~GeV, at LO and NLO, (a) as function of c.m.s. pseudorapidity $\eta$ at 
fixed $P_{h\perp}=3$~GeV, (b) as function of $P_{h\perp}$, integrated over $|\eta|\leq 2$.}
\end{figure*}

Interestingly, NLO corrections to the asymmetry $A_{LL}$ become quite important for a smaller fixed transverse hadron momentum $P_{h\perp}=3$~GeV. In this case the event rate is about 200 times larger compared to the one at a large transverse momentum $P_{h\perp}=10\,\mathrm{GeV}$. In Fig.~\ref{fig:EICeta} we plot the asymmetry $A_{\mathrm{LL}}$ for $e p\to \pi^+X$ at the EIC as a function of the pion's c.m.s. pseudorapidity $\eta$, at a fixed pion transverse momentum $P_{h\perp}=3$~GeV. 
 At midrapidity (where the event rate is largest) the asymmetry is about $2\%$. The asymmetry is considerably affected by NLO-corrections for pseudorapidities $\eta > 0.5$. In this region we observe a $60\%$ reduction of the asymmetry when going from LO to NLO. This effect 
is generated by large $K$-factors of the unpolarized NLO cross section, caused by dominant real photon contributions. The unpolarized cross section receives positive enhancements from all partonic channels. On the other hand, the spin-dependent cross section obtains a relatively large negative contribution from the gluon-induced subprocess which partly compensates the large positive enhancements from the quark induced channels. Overall, this leads to an NLO correction that is smaller for the spin-dependent cross section than for the spin-averaged one, and consequently to a large NLO effect on the asymmetry. This sensitivity to gluon-induced processes at NLO indicates an opportunity to constrain the gluon's helicity distribution $\Delta g$ at the EIC.

The $P_{h\perp}$-dependence of the spin asymmetry at the EIC is shown in Fig.~\ref{fig:EICpt}.
As expected, the asymmetry becomes smaller at lower pion transverse momenta and the NLO correction lowers the asymmetry somewhat.

%%%%%%%%%%%%%%%%%%%%%%%%%%%%%%%%%%%%%%%%%%%%%%%%%%%%%%%%%%%%
%%%%%%%%%%%%%%%%%%%%%%%%%%%%%%%%%%%%%%%%%%%%%%%%%%%%%%%%%%%%

%%%%%%%%%%%%%%%%%%%%%%%%%%%%%%%%%%%%%%%%%%%%%%%%%%%%%%%%%%
\begin{figure*}[t]
\centering
\subfloat[]{\includegraphics[width=0.5\textwidth,angle=0]{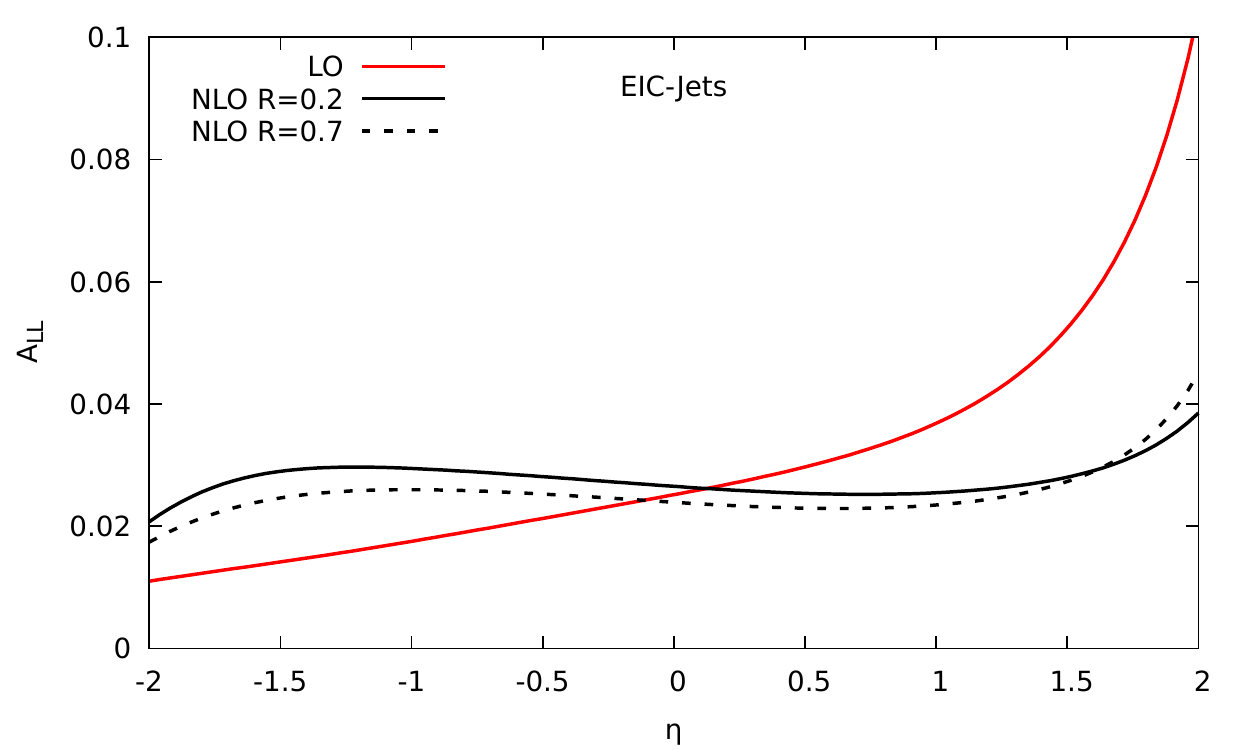}\label{fig:jets}}
\subfloat[]{\includegraphics[width=0.5\textwidth,angle=0]{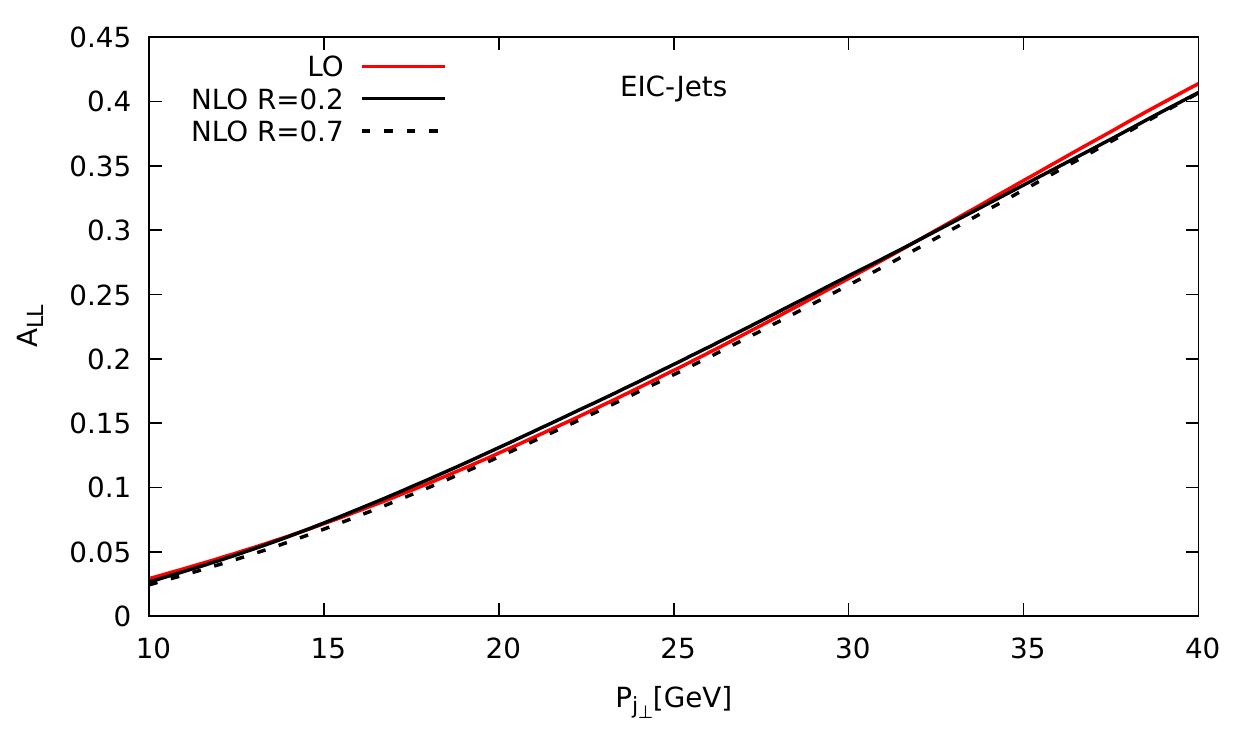}\label{fig:jets1}}
\caption{Longitudinal double-spin asymmetry for single-inclusive jet production at the EIC, (a) as function of 
jet pseudorapidity $\eta$ at a fixed transverse jet momentum $P_{j\perp}=10$~GeV, and (b) as function of 
$P_{j\perp}$, integrated over $|\eta|\leq 2$. We have used the anti-$k_t$ jet algorithm~\cite{Cacciari:2008gp}.
We show NLO predictions for two different values of the jet size parameter, $R=0.7$ and $R=0.2$.}
\end{figure*}
%%%%%%%%%%%%%%%%%%%%%%%%%%%%%%%%%%%%%%%%%%%%%%%%%%%%%%%%%%

Given the high energy of an EIC, also jet observables will become available. Therefore, we also present predictions 
for the double-longitudinal spin asymmetry $A_{\mathrm{LL}}$ in $ep\to\mathrm{jet}\,X$, using the NLO calculations
described in Sec.~\ref{jetpro}. We adopt the anti-$k_t$ jet algorithm of~\cite{Cacciari:2008gp}. 
Figure~\ref{fig:jets} shows the dependence of $A_{\mathrm{LL}}$ on the jet's pseudorapidity $\eta$ at fixed jet transverse 
momentum $P_{j\perp}=10\,\mathrm{GeV}$. We plot the NLO asymmetry for two jet sizes, $R=0.2$ and $R=0.7$.
For negative jet-pseudorapidities the NLO corrections seem to increase the asymmetry with 
respect to the LO result, while they decrease it for positive pseudorapidities. The $P_{j\perp}$-dependence of the asymmetry is 
shown in Fig.~\ref{fig:jets1}. It turns out to remain largely unaffected by the NLO corrections.

\section{Conclusions\label{concl}}

We have performed next-to-leading order calculations of the spin-dependent partonic cross sections for the processes
$\ell N\to h X$ and $\ell N\to \mathrm{jet}\,X$ for longitudinal polarization of the initial particles. Based on these
results we have computed the double-longitudinal spin asymmetry $A_{\mathrm{LL}}$ to NLO accuracy. We have
found that the NLO corrections tend to reduce the size of the spin asymmetry.  

We have presented detailed comparisons of our $A_{\mathrm{LL}}$ to the data by the SLAC E155 experiment
which has measured the asymmetry for $ep$ and $eD$ scattering, in each case both for charged pions and for
unidentified charged hadrons $h^\pm$. Data were recorded separately for two scattering angles, 
$\theta=2.75^\circ$ and $\theta=5.5^\circ$. 
No consistent picture emerges from the comparisons. By and large, the theoretical asymmetry lies 
higher than the data. For scattering off a deuteron target there is typically at least a qualitative agreement
between the NLO calculation and the data. A notable exception is the asymmetry for $eD\to h^+X$. 
For scattering off a proton target, some of the asymmetries are very badly described, with the theoretical results 
being much higher than the data. This is the case especially for the asymmetries for $ep\to \pi^+X$ at the lower 
scattering angle and for $ep\to h^+X$ for both angles.

It is difficult to draw clear-cut conclusions from these findings. It is possible that non-perturbative power-suppressed 
contributions are still relevant in kinematic regimes relevant for E155, which would invalidate the use
of QCD perturbation theory. Assuming that this is not the case, one question concerns the role of 
QCD corrections beyond NLO. As we have seen, the asymmetries decrease when going from LO
to NLO so that it is conceivable that this trend will continue when even higher orders are taken into
account. While a NNLO calculation of the spin-averaged cross section has now been carried out for
$\ell N\to {\mathrm{jet}}\,X$~\cite{JetNNLO}, no such calculation exists presently for $\ell N\to hX$ or for
the double-spin asymmetry. On the other hand, it may well be that the bulk of the beyond-NLO corrections
can be estimated using QCD threshold resummation techniques. A related study has recently been
performed for the process $\ell N\to \ell' h X$ in photoproduction (that is, with an observed final-state
lepton)~\cite{Uebler}, and it was indeed found that the higher-order corrections further suppress
the asymmetry. However, this suppression will likely not be significant enough to bridge the partly
large differences between data and theory we find. 

Arguably the physically most interesting explanation for the observed discrepancies would resort
to changes in the helicity parton distributions. In this context it is interesting to note that recent data for 
the spin asymmetry $A_{\mathrm{LL}}$ in photoproduction via $\mu N\to\mu' h X$ published by
COMPASS~\cite{Adolph:2015hta} also show a trend that the deuteron asymmetry is better described 
by theory than the proton one. On the other hand, unlike for photoproduction,
for $\ell N\to hX$ only quarks participate at
the lowest order. For the kinematics relevant at E155 the average
values of the incoming parton's momentum fraction $x$ are relatively large, so that the up and down 
valence helicity distributions make the dominant contributions to the spin-dependent cross section. 
These distributions are rather well constrained, so there is little room for major changes here. The
gluon helicity distribution is still known with much poorer accuracy; however, again in contrast to
photoproduction, gluonic channels enter only at NLO and hence the sensitivity of $A_{\mathrm{LL}}$ 
to $\Delta g$ is relatively weak. 
Overall it appears unlikely that the discrepancies between data and theory that we observe 
are due to the helicity parton distributions alone, especially given that the NLO proton asymmetries for
E155 kinematics would need to decrease very strongly (see for example Figs.~\ref{fig:ALLp2hp275}
and~\ref{fig:ALLp2hp55}). Clearly, further studies are needed here. 

We hope that other experiments can obtain new data for $A_{\mathrm{LL}}$ in single-inclusive 
lepton scattering. We thus have presented predictions for the spin asymmetry for HERMES, JLab12, COMPASS
and the electron ion collider. We expect that latter to provide particularly valuable information. 
Data, if available with sufficient precision and large lever arm in 
kinematics, might help to clarify whether and when the process can be reliably described by perturbative QCD. 
As discussed in the Introduction, this would in turn have important ramifications also for our understanding of 
single-transverse spin asymmetries, since a proper understanding of the simpler leading-twist observables in
these single-inclusive processes is required before one can reliably address the more complicated transverse spin effects.

%%%%%%%%%%%%%%%%%%%%%%%%%%%%%%%%%%%%%%%%%%%%%%%%%%%%%%%%%
\begin{acknowledgments}
We thank A.~Metz for interesting discussions that have initiated this project, and E.~Aschenauer and M.~Stratmann for
helpful comments. This work was supported by the ``Bundesministerium f\"{u}r Bildung und Forschung'' 
(BMBF) grant 05P15VTCA1. 

\end{acknowledgments}
%%%%%%%%%%%%%%%%%%%%%%%%%%%%%%%%%%%%%%%%%%%%%%%%%%%%%%%%%

\appendix
\section{NLO coefficients} \label{App:AppendixA}

Here we present the NLO coefficients in Eqs.~(\ref{Resq2qNLOreal1}), (\ref{Resq2gNLOreal1}), (\ref{Resg2qNLOreal1}) 
for the various partonic channels. We start with inclusive-hadron production:

\paragraph{$q\to q$ channel:}
\begin{eqnarray}
\Delta A_0^{q\to q} & = & \frac{1+v}{1-v}
\left((3 +2\log v) \log\left(\frac{s(1-v)}{\mu^2}\right)+\log^2 v-8\right),\nonumber\\[2mm]
\Delta A_{1}^{q\to q} & = & 8 w \frac{1-v(1-2w)}{1-v},\nonumber\\[2mm]
\Delta B_{1}^{q\to q} & = & 4w\frac{1+vw}{1-v},\nonumber\\[2mm]
\Delta B_{2}^{q\to q} & = & 4w\frac{1-v(1-2w)}{1-v},\nonumber\\[2mm]
\Delta B_{3}^{q\to q} & = & 4 w \frac{1-v(1-2w)}{1-v},
\end{eqnarray}
\begin{eqnarray}
\Delta C_{1}^{q\to q}& = & \frac{1}{(1-v)(1-v(1-w))}\Big[2-w+vw(9-w)\nonumber\\
&&-v^2(2+2w-3w^2)-2v^3w(1-3w+2w^2)\Big],\nonumber\\[2mm]
\Delta C_{2}^{q\to q} & = & \frac{2(1+v)}{1-v},\nonumber\\[2mm]
\Delta C_{3}^{q\to q}& = & -\frac{2vw(1-v(1-2w))}{1-v},\nonumber\\[2mm]
\Delta C_{4}^{q\to q}& = & \frac{2v^2w(1-w)(1-v(1-2w))}{(1-v)(1-v(1-w))},\nonumber\\[2mm]
\Delta C_{5}^{q\to q} & = & \frac{1}{(1-v)(1-v(1-w))}\Big[2-w+vw(5-w)\nonumber\\
&&-v^2(2-2w+w^2)-2v^3w(1-3w+2w^2)\Big],\nonumber\\[2mm]
\Delta C_{6}^{q\to q} & = & \frac{(1-v)(1-w)(1+vw)}{1-v(1-w)}.\label{Cq2q}
\end{eqnarray}
\\
\\
\paragraph{$q \to g$ channel:}

\begin{eqnarray}
\Delta C_{1}^{q\to g}& = & \frac{2vw(1-v(1-2w))(1+v^2(1-w)^2)}{(1-v)(1-v(1-w))^2},\nn\\[2mm]
\Delta C_{2}^{q\to g}& = & -\frac{2vw(1+v(1-2w))}{1-v},\nonumber\\[2mm]
\Delta C_{3}^{q\to g} & = & \frac{vw}{(1-v)(1-vw)^2(1-v(1-w))^2}\Big[1-v(1-2w)\nonumber\\
&&+v^2(1-2w-3w^2)-v^3(1-7w^2+4w^3)\nonumber\\
&&+2v^4w(2-7w+7w^2-2w^3)\nonumber\\
&&-2v^5w^2(1-w)^2(1-2w)\Big],\nonumber\\[2mm]
\Delta C_{4}^{q\to g} & = & \frac{vw}{(1-v)(1-vw)^2(1-v(1-w))^2}\Big[2-4v\nonumber\\
&&+v^2(4-13w+12w^2)-v^3(2-14w+13w^2)\nonumber\\
&&-v^4w(3-4w+3w^2-2w^3)+v^5w^2(1-w)^2\Big].\label{Cq2g}
\end{eqnarray}

\paragraph{$g \to q$ channel:}

\begin{eqnarray}
\Delta C_{1}^{g\to q}& = & -\frac{2(1+v(1-2w))}{1-v},\nonumber\\[2mm]
\Delta C_{2}^{g\to q} & = & -\frac{1}{(1-v)(1-vw)^2}\Big[2-2w+2v(1-4w+2w^2)\nonumber\\
&&-v^2w(4-11w+2w^2)+3v^3w^2(1-2w)\Big],\nonumber\\[2mm]
\Delta C_{3}^{g\to q} & = & \frac{1}{(1-v)(1-vw)^2}\Big[2(1-w)+2v(2-5w+2w^2)\nonumber\\
&&-v^2w(7-12w+2w^2)+v^3w^2(7-8w)\Big].\label{Cg2q}
\end{eqnarray}

For the single-inclusive jet cross section in Eq.~(\ref{jetPartCS}) we have the following
coefficients:
\begin{eqnarray}
A_0^{\mathrm{jet}}&=&\left(\frac{3}{2}+2\log v\right)\,\log\left(\frac{v(1-v)s}{\mu^2}R^2\right)+2\log^2 v-
\frac{13}{2}+\frac{2}{3}\pi^2,\nonumber\\
A_1^{\mathrm{jet}}&=&4w,\nonumber\\
B_1^{\mathrm{jet}}&=&2w\,\log\left(\frac{wv^3(1-v)s}{\mu^2}R^2\right),\nonumber\\
C_1^{\mathrm{jet}}&=&\frac{vw}{1-v(1-w)}\times\nonumber\\
&&\left[\frac{1+v(1-w)}{1-v(1-w)}\,\log\left(\frac{w(1-w)^2v^3(1-v)s}{\mu^2}R^2\right)+1\right],\label{Cjet}
\end{eqnarray}
where $R$ is the jet size parameter. 
The coefficient $A_0^{\mathrm{jet}}$ depends on the jet algorithm adopted \cite{Mukherjee:2012uz}. 
The result given above applies to the anti-$k_T$ algorithm.

%%%%%%%%%%%%%%%%%%%%%%%%%%%%%%%%%%%%%%%%%%%%%%%%%%%%%%%%
%%%%%%%%%%%%%%%%%%%%%%%%%%%%%%%%%    References    %%%%%%%%%%%%%%%
%%%%%%%%%%%%%%%%%%%%%%%%%%%%%%%%%%%%%%%%%%%%%%%%%%%%%%%%

\end{document}